\documentclass[10pt,a4paper]{article}            % DO NOT DELETE THIS LINE
\usepackage{geometry}                % See geometry.pdf to learn the layout options. There are lots.
\usepackage{fancyhdr}
\usepackage{authblk}
\usepackage{graphicx}
\usepackage{url}
\makeatletter
\def\url@leostyle{%
  \@ifundefined{selectfont}{\def\UrlFont{\sf}}{\def\UrlFont{\small\ttfamily}}}
\makeatother
\usepackage{epstopdf}
\usepackage{amssymb}
\usepackage[fleqn]{amsmath}
\def\mathbi#1{\textbf{\em #1}}
\numberwithin{equation}{section}
\DeclareMathSymbol{\Gamma}{\mathalpha}{letters}{"00}
\DeclareMathSymbol{\Lambda}{\mathalpha}{letters}{"03}
\DeclareMathSymbol{\Omega}{\mathalpha}{letters}{"0A}
\DeclareMathAlphabet{\mathitbf}{OML}{cmm}{b}{it}

\begin{document}                  % DO NOT DELETE THIS LINE

     %-------------------------------------------------------------------------
     % The introductory (header) part of the paper
     %-------------------------------------------------------------------------

     % The title of the paper. Use \shorttitle to indicate an abbreviated title
     % for use in running heads (you will need to uncomment it).

\title{The Geometry of Niggli Reduction I:  The Boundary Polytopes of the Niggli Cone}
%\shorttitle{Short Title}

     % Authors' names and addresses. Use \cauthor for the main (contact) author.
     % Use \author for all other authors. Use \aff for authors' affiliations.
     % Use lower-case letters in square brackets to link authors to their
     % affiliations; if there is only one affiliation address, remove the [a].

\author[1]{Lawrence C. Andrews }
\affil[1]{Micro Encoder  Inc., 11533 NE 118th St, \#200,  Kirkland, WA 98034-7111 USA}
\author[2,*]{Herbert J. Bernstein }
\affil[2]{Dowling College, 1300 William Floyd Parkway, Shirley, NY 11967 USA}
\affil[*]{To whom correspondence should be addressed. Email: {\it yaya@dowling.edu}}
%\author{Lawrence C. Andrews//Micro Encoder  Inc., 11533 NE 118th St, \#200,  Kirkland, WA 98034-7111 USA /and Herbert J. Bernstein //Dowling College, 1300 William Floyd Parkway, Shirley, NY 11967 USA yaya@dowling.edu}
%\author[b]{}
%affil[a]{Micro Encoder  Inc., 11533 NE 118th St, \#200,  Kirkland, WA 98034-7111 \country{USA}}
%\affil[b]{Dowling College, 1300 William Floyd Parkway, Shirley, NY 11967 \country{USA}}

     % Use \shortauthor to indicate an abbreviated author list for use in
     % running heads (you will need to uncomment it).

%\shortauthor{Andrews and Bernstein}

     % Use \vita if required to give biographical details (for authors of
     % invited review papers only). Uncomment it.

%\vita{Author's biography}

     % Keywords (required for Journal of Synchrotron Radiation only)
     % Use the \keyword macro for each word or phrase, e.g. 
     % \keyword{X-ray diffraction}\keyword{muscle}

%\keyword{keyword}

     % PDB and NDB reference codes for structures referenced in the article and
     % deposited with the Protein Data Bank and Nucleic Acids Database (Acta
     % Crystallographica Section D). Repeat for each separate structure e.g
     % \PDBref[dethiobiotin synthetase]{1byi} \NDBref[d(G$_4$CGC$_4$)]{ad0002}

%\PDBref[optional name]{refcode}
%\NDBref[optional name]{refcode}

\maketitle                        % DO NOT DELETE THIS LINE

%\begin{synopsis}
%This is an investigation of the boundary polytopes of the Niggli-reduced cone ${\mathbi{N}}$ in the six-dimensional space ${\mathbi{G}^{\mathbi{6}}}$ by 
%a combination of algebraic analysis of the Niggli reduction transformations and by algebraic analysis and organized random probing of regions near 1-, 2-, 3-, 4-, 5-, 6-, 7- and 8-fold %boundary polytopes.  We limit our consideration of valid boundary polytopes to those avoiding the mathematically interesting but crystallographically impossible cases of zero length cell %edges.   Combinations of boundary polytopes without a valid intersection in the closure of the Niggli cone or with an intersection that would force a cell axis to zero or without neighboring %probe points are eliminated.  216 boundary polytopes are found.    
%15 5-D boundary polytopes provide a new, simpler and arguably more intuitive basis set for the classification of lattice characters.  The
%classification is intended to help in organizing database searches and in understanding which lattice symmetries are ``close'' to a given
%experimentally determined cell.
%\end{synopsis}

\begin{abstract}
Correct identification of the Bravais lattice of a crystal is an important early step in structure solution. Niggli reduction is a commonly used technique. We investigate the boundary polytopes of the Niggli-reduced cone ${\mathbi{N}}$ in the six-dimensional space ${\mathbi{G}^{\mathbi{6}}}$ by algebraic analysis and organized random probing of regions near 1-, 2-, 3-, 4-, 5-, 6-, 7- and 8-fold boundary polytope intersections.  We limit our consideration of valid boundary polytopes to those avoiding the mathematically interesting but crystallographically impossible cases of zero length cell edges.   Combinations of boundary polytopes without a valid intersection in the closure of the Niggli cone or with an intersection that would force a cell edge to zero or without neighboring probe points are eliminated.  216 boundary polytopes are found.  There are
15 5-D boundary polytopes of the full ${\mathbi{G}^{\mathbi{6}}}$ Niggli cone ${\mathbi{N}}$,  53 4-D boundary polytopes resulting from intersections of pairs of the 15 5-D boundary polytopes, 79 3-D boundary polytopes resulting from 2-fold, 3-fold and 4-fold intersections of  the 15 5-D boundary polytopes, 55 2-D boundary polytopes resulting from 2-fold, 3-fold, 4-fold and higher intersections of the 15 5-D boundary polytopes, 14 1-D boundary polytopes resulting from 3-fold and higher intersections of the 15 5-D  boundary polytopes. 

All of the primitive lattice types can be represented as combinations of the 15 5-D boundary polytopes.  
All of the non-primitive lattice types can be represented as combinations of
the 15 5-D boundary polytopes and of the 7 special-position subspaces of the 5-D boundary polytopes.
This study provides a new, simpler and arguably more intuitive basis set for the classification
of lattice characters and helps to illuminate some of the complexities in Bravais lattice identification.
The classification is intended to help in organizing database searches and in understanding which 
lattice symmetries are ``close'' to a given experimentally determined cell.

\end{abstract}

     %-------------------------------------------------------------------------
     % The main body of the paper
     %-------------------------------------------------------------------------
     % Now enter the text of the document in multiple \section's, \subsection's
     % and \subsubsection's as required.
     
\section{Introduction}

This is the first in a series of papers on understanding and classifying crystallographic
lattices and the cell parameters commonly used to identify those lattices.  This paper
is concerned with fundamental mathematical issues underlying the subsequent papers.
The next two papers are concerned with the application of this approach to lattice identification
\cite{Andrews2013}
and to unit cell databases \cite{McGill2013}.  More papers based on this formalism are in preparation.
The boundary and matrix definitions from this paper
are needed for a full understanding of the subsequent papers. 

The wide-spread use of Niggli reduction in crystallography implies that it
should be thoroughly understood.  Robust identification of Bravais
lattices and lookup of unit cell parameters in databases would be improved
if a successful embedding of the Niggli space could be achieved.  An essential
preliminary step towards doing such an embedding is the mapping of the
boundaries of the space of Niggli-reduced cells.

Failure to correctly identify the Bravais lattice of a crystal can compromise subsequent least-squares 
calculations or even the solution of a structure.     
There are two commonly used ways to obtain a unique representative of the infinite
number of cells that may be used to generate a given lattice:  Niggli reduction \cite{Niggli1928} and
Delaunay reduction \cite{Delaunay1932}.  We follow the conventions of the International Tables for Crystallography
 \cite{Burzlaff1992} in basing this paper on Niggli reduction, and we will recast the discussion in terms of
Delaunay reduction in a future study.  
Niggli reduction defines a complex space that has not previously been fully analyzed. 
Several authors have published interesting commentaries on the properties of this space \cite{Hosoya2000} \cite{Oishi-Tomiyasu2012}  \cite{Gruber1997}. These studies use the space ${\mathbi{G}^{\mathbi{6}}}$ \cite{Andrews1988}, or a similar metric-tensor-based space, or 
a projection of ${\mathbi{G}^{\mathbi{6}}}$ to a space of lower dimensionality, respectively. Two principal uses of Niggli reduction are 
the determination of Bravais lattice type and the construction of databases using a representation of the unit cell for the key \cite{Andrews1980} \cite{Toby1994} \cite{Byram1996}.

Given a precisely determined reduced cell, the lattice symmetry may be unambiguously inferred.  In addition, reduced cells
are useful in searching for sub- and super-cells, indexing of powder patterns and in  twinned crystal studies.
Without a suitable metric and a clear understanding of the geometry of the boundaries
of the space, searches of lattices in the neighborhood of a given cell must either be excessively broad and
produce many false positives, or, if made tighter, risk missing important candidates, especially in the vicinity
of 90 degree angles.

As noted in \cite{Azaroff1958} the concept of a reduced cell is strongly related to the  concept of a reduced 
ternary quadratic form \cite{Seeber1831} \cite{Selling1874}.  Reduced cells have become an important
computational tool in crystallography (just as reduced quadratic forms are an important tool in
computational number theory) but much of the literature focuses on an essentially qualitative
approach, taking a reduced cell as an absolute indicator of symmetry and not considering the impact
of experimental error \cite{Andrews1988}.  Consideration of experimental error requires use of a distance metric
dealing with reduced cells.   Such metrics have been used in a 6-dimensional approach \cite{Zimmermann1985}, another
6-dimensional approach  \cite{Andrews1988},
a B-matrix approach \cite{Macicek1992}  and a ``distortion index'' approach \cite{Minor1997}.

In addition, because the processes of both
Niggli and Delaunay reduction can produce large discontinuities in reduced cell parameters from small changes in the lattice, 
an effective use of a metric must allow for such discontinuities, either by a combinatorial search or by a metric-preserving
embedding in a higher dimensional space that removes the discontinuities.  Until now, crystallographic software
appears either to have used a demanding combinatorial approach or simply to have given up on a doing complete search.   
The purpose of this paper is to take the necessary first steps towards adding an embedding to the existing
combinatorial approaches by clearly mapping the boundary discontinuities and their related transformations.

There have been multiple investigations of such boundaries, albeit  without a consideration of experimental error in most cases.  
See, for example, Gruber \cite{Gruber1997} \cite{Gruber2006} for a review and an approach using 
a 5-dimensional space based on the metric tensor.  Gruber's 1997 approach partitions the space of 
reduced  cells into 127 disjoint components (genera), based on 67 1- to 4-dimensional ``hyperfaces'' further subdivided
into  227 hyperfaces
in order to achieve a common partitioning applicable to both
Niggli and Delaunay reduction.   Unfortunately,  Gruber's reduction to 5 dimensions,
and any purely topological approach without a metric that allows error propagation from the experimental
data,  makes it difficult or
impossible to do a full perturbation analysis of the impact of experimental errors.

Therefore in this investigation we return to the full 6-dimensional space \cite{Andrews1988}, 
${\mathbi{G}^{\mathbi{6}}}$, of unit cells based on the metric tensor and use algebraic techniques confirmed by 
a Monte Carlo technique to explore the 
``natural''  5-, 4-, 3-, 2- and 1-dimensional boundary polytopes of the 6-dimensional polytope of Niggli-reduced cells.
The two techniques are mutual supportive. The lower dimensional boundaries are an algebraic consequence of the 5-dimensional boundaries.
The lower dimensional boundaries derived algebraically are confirmed by the Monte Carlo technique
which helps to identify unpopulated boundaries and boundaries that drop to lower dimension due to glancing
intersections with multiple other boundaries.

There are 15 5-dimensional boundary polytopes and a total of 216 boundary polytopes.  
In this approach, all the boundary polytopes are on the surface of the closure of the 6-dimensional Niggli-reduced polytope.
Identification of these boundary polytopes and the reduction transformations to be applied in crossing them is
an essential step in either a combinatorial error analysis or in embedding ${\mathbi{G}^{\mathbi{6}}}$ into a higher dimensional space
for an analytic error analysis, and it is useful for database searches.  

The 15 5-D boundary polytopes give the complete shape of the space of Niggli-reduced cells.
All of the primitive lattice types can be represented as combinations of the 15 5-D boundary polytopes.  
All of the non-primitive lattice types can be represented as combinations of
the 15 5-D boundary polytopes and of the 7 special-position subspaces of the 5-D boundary polytopes.
By confining our attention to just the Niggli reduction, the result is a simpler classification than Gruber's
with more direct applicability to an embedding and database searching.  The embedding and
database searches are the subjects of  following studies.

\section{The space ${\mathbi{G}^{\mathbi{6}}}$}

${\mathbi{G}^{\mathbi{6}}}$ is a reformulation of the crystallographic metric tensor and the ``Niggli matrix'' (itself a reformulation of the metric tensor) \cite{Andrews1988}. A vector $\overrightarrow{g}$ in ${\mathbi{G}^{\mathbi{6}}}$ is defined as:
\[
\overrightarrow{g} =\left(~~ \overrightarrow{a}\!\!\cdot\!\overrightarrow{a} ,
~~\overrightarrow{b}\!\!\cdot\!\overrightarrow{b} ,
~~\overrightarrow{c}\!\!\cdot\!\overrightarrow{c} ,
~~2 \overrightarrow{b}\!\!\cdot\!\overrightarrow{c} ,
~~2 \overrightarrow{a}\!\!\cdot\!\overrightarrow{c} ,
~~2 \overrightarrow{a}\!\!\cdot\!\overrightarrow{b}  \right)
\]
\[
= (~~ \| \overrightarrow{a} \| ^2,
~~ \| \overrightarrow{b} \| ^2,
~~ \| \overrightarrow{c} \| ^2,
~~ 2 \| \overrightarrow{b} \| \| \overrightarrow{c} \| cos (\alpha),\]
\[
 ~~~~~~~~~~
~~ 2 \| \overrightarrow{a} \| \| \overrightarrow{c} \| cos (\beta),
~~ 2 \| \overrightarrow{a} \| \| \overrightarrow{b} \| cos (\gamma) )
\]
\[
= \left( g_1,
~~ g_2,
~~ g_3,
~~ g_4,
~~ g_5,
~~ g_6 \right)
= g_{\{ 1, 2, 3\}},  g_{\{ 4, 5, 6\}}
\]
where $\overrightarrow{a}$, $\overrightarrow{b}$, $\overrightarrow{c}$ are the 
unit cell edge vectors, and ``$\cdot$'' indicates the dot product.  The unit cell 
is chosen to be primitive.

The notation $g_{\{ 4, 5, 6\}}$ is used to denote the elements $ \left(g_4,g_5, g_6 \right)$ 
from the full ${\mathbi{G}^{\mathbi{6}}}$ vector.

\section{The Niggli Conditions}

The Niggli-reduced cell of a lattice is a unique choice from among the infinite number 
of alternate cells that generate the same lattice \cite{Niggli1928}.  A Buerger-reduced cell 
for a given lattice is any cell that generates that lattice, chosen such that no other cell 
has shorter cell edges \cite{Buerger1960}.   Even after allowing for the equivalence of 
cells in which the directions of axes are reversed or  axes of the same length are 
exchanged, there can be up to five alternate Buerger-reduced cells for the same 
lattice \cite{Gruber1973}.  The Niggli conditions allow the selection of a unique
reduced cell for a given lattice from among the alternate Buerger reduced cells for that lattice.

Niggli reduction consists of converting the original cell to a primitive one 
and then alternately applying two operations: 
conversion to standard presentation and reduction \cite{Niggli1928} \cite[Table 1]{Andrews1988}.  The 
convention for meeting the combined Buerger and Niggli conditions is based on 
increasingly restrictive layers of constraints:

If $g_1 < g_2 < g_3$,  $|g_4| < g_2$, $|g_5| < g_1$, $|g_6| < g_1$ and 
either $g_{\{ 4, 5, 6\}} > 0$ or  $g_{\{ 4, 5, 6\}} \leqslant 0$ then we have 
a Niggli-reduced cell, and we are done. 

The remaining conditions are imposed when any of the above inequalities becomes 
an equality or the elements of 
$g_{\{ 4, 5, 6\}}$ are not consistently all strictly positive or are not consistently 
all less than or equal to zero.

The full set of combined Buerger and Niggli conditions in addition to those for the cell edge lengths being minimal is:
\[ \text{require}\ 0 \leqslant g_1 \leqslant g_2 \leqslant g_3 \]
\begin{equation}\text{if} \ g_1 = g_2, \text{then require}\  |g_4| \leqslant |g_5| \label{celleq12}\end{equation}
\begin{equation}\text{if}\  g_2 = g_3, \text{then require}\  |g_5| \leqslant |g_6| \label{celleq23}\end{equation}
\[ \text{require} \  \{g_4 > 0 \  \text{and}\   g5> 0 \   \text{and}\   g6 > 0\}\  \]
\[ \text{or  require}\  \{g_4 \leqslant 0 \  \text{and}\  g_5 \leqslant 0 \ \text{and}\   g6 \leqslant 0\} \]
\[ \text{require}\ |g_4| \leqslant g_2 \]
\[ \text{require}\ |g_5| \leqslant g_1 \]
\[ \text{require}\ |g_6| \leqslant g_1 \]
\[ \text{require}\ g_3 \leqslant g_1+ g_2 + g_3 + g_4 + g_5 +g_6\]
\begin{equation}\text{if} \  g_4 = g_2, \text{then require}\  g_6 \leqslant 2 g_5 \label{celleq42}\end{equation}
\[\text{if} \  g_5 = g_1,  \text{then require}\  g_6 \leqslant 2 g_4\]
\[\text{if} \  g_6 = g_1, \text{then require}\  g_5 \leqslant 2 g_4\]
\[\text{if} \  g_4 = -g_2, \text{then require}\  g_6 = 0\]
\[\text{if} \  g_5 = -g_1, \text{then require}\  g_6 = 0\]
\begin{equation}\text{if} \  g_6 = -g_1, \text{then require}\  g_5 = 0 \label{celleq6m1}\end{equation}
\[ \text{if} \  g_3 =  g_1+ g_2 + g_3 + g_4 + g_5 +g_6,  \text{then require}\ 2 g_1 + 2 g_5 + g_6 \leqslant 0 
\nonumber\]

\noindent{}The ${\mathbi{G}^{\mathbi{6}}}$ transformations associated with each of these steps are enumerated in \cite{Andrews1988}.  Application of these operations must be repeated until all are satisfied.

\section{Notation and Boundary Polytopes}

We define the manifold of Niggli reduced cells in ${\mathbi{G}^{\mathbi{6}}}$ as ${\mathbi{N}}$ and refer to it as the ``Niggli cone''.

The interior of ${\mathbi{N}}$, $int({\mathbi{N}})$, is defined as the set of $\overrightarrow{n} \in N$ such that
there exists $r > 0$ such that for all $\overrightarrow{g} \in {\mathbi{G}^{6}}$ such that $\| \overrightarrow{g}  - \overrightarrow{n} \|  < r,
\overrightarrow{g} \in N$.

The closure of ${\mathbi{N}}$, $cl({\mathbi{N}})$, is defined as the set of $\overrightarrow{g} \in {\mathbi{G}^{6}}$ such that
for all $r > 0$ there exists $\overrightarrow{n} \in {\mathbi{N}}$ such that $\| \overrightarrow{g} - \overrightarrow{n} \|  \leqslant r$.

The boundary of ${\mathbi{N}}$, $\partial ({\mathbi{N}})$, is defined as the set of points in  $cl({\mathbi{N}})$ not in $int({\mathbi{N}})$.

The boundary of ${\mathbi{N}}$ is created by the linear constraints of Niggli reduction and therefore can be
decomposed into the union of ``polytopes,'' {\it i.e.} flat facets with straight edges.

We distinguish the primary boundary polytopes from their edges, which are also polytopes, by
``dimension'' which is the number of vectors in a basis for the interior of the polytope.  

${\mathbi{N}}$ is a six-dimensional (6-D) polytope.  ${\mathbi{N}}$ is a double-ended cone-like region going through
 the origin to infinity in both directions.  The 
boundary  polytopes are flat facets created by the intersections of hyperplanes through the origin.
The boundary polytopes are, of course, of lower-dimension than ${\mathbi{N}}$.  Therefore any randomly selected
vector in ${\mathbi{G}^{\mathbi{6}}}$ has a vanishingly small probability of occupying any particular 5-dimensional (``5-D'')
boundary polytope, and it has an even lower probability of occupying one of the lower-dimensional boundary polytopes resulting from the intersections of 5-D boundary polytopes.
Some boundary polytopes are ``\textbf{open}'', \textit{i.e.} while there are Niggli-reduced cells near that boundary, some or all of the points
on those boundary polytopes are not themselves Niggli-reduced.

Our task is to identify the 5-D boundary polytopes that give ${\mathbi{N}}$ its shape.  Those 5-D boundaries
and the transforms involved in crossing them generate all the rest of the structure.  However, in order to understand
the shape of a given 5-D boundary polytope, we need the 4-D edges that bound it.
In order to understand
the shape of a given 4-D boundary polytope, we need the 3-D edges that bound it.  
In order to understand
the shape of a given 3-D boundary polytope, we need the 2-D edges that bound it.  
In order to understand
the shape of a given 2-D boundary polytope, we need the 1-D edges that bound it.  
From this classification we gain a better understanding
of the relationships among Bravais lattice types, and, perhaps more importantly, this provides 
essential information needed to organize computations.   Hosoya \cite{Hosoya1990} addressed a different, but related,
classification.  We discuss the relationship between boundary polytope identification, lattice types and Hosoya's approach in 
section \ref{sec:rel}.  Hosoya recognized the complexity of boundary identifications for ${\mathbi{N}}$ and introduced the
use of Monte Carlo searching to clarify the relationships.  We apply a Monte Carlo search in Appendix \ref{sec:mc}.

The reduction steps convert a non-reduced cell into one that has at least one 
edge shorter than the starting edges, and other steps in the case of equality 
convert a non-reduced cell into a cell that is more orthogonal than the starting cell. 
These operations are accomplished by choosing a face- or body-diagonal to replace 
one of the cell edges.  The conditions added to remove the ambiguities in the case 
of equalities allow for a unique choice of Niggli cell in all cases but thereby create 
complex boundary conditions \cite{Andrews1988}.  For example, the cell edge equalities in 
equations (\ref{celleq12}) and (\ref{celleq23}) create boundary polytopes across 
which elements of both $g_{\{1,2,3\}}$ and of $g_{\{4,5,6\}}$ are exchanged, while 
equations (\ref{celleq42}) through (\ref{celleq6m1}) create boundary polytopes across 
which edges are exchanged for face-diagonals.  
 
A boundary polytope does not necessarily consist entirely of Niggli-reduced cells, but must be 
``near'' Niggli-reduced cells and must be ``near'' non-Niggli-reduced cells as well.   
 We define a boundary polytope, ${\mathitbf{\Gamma}} \subseteq 
\!{\mathbi{N}}$ as a subset of ${\it \mathbi{N}}$ for which there is an associated matrix $M_{{\mathitbf{\Gamma}}}$
such that for all points $\overrightarrow{\gamma} \in {\mathitbf{\Gamma}}$ there exists $\delta > 0$ such that for all $\overrightarrow{nn} \notin {\mathbi{N}}$ where
$\|\overrightarrow{nn} - \overrightarrow{\gamma}\| < \delta,\    M_{{\mathitbf{\Gamma}}}(\overrightarrow{\gamma}) \in \mathbi{N}$.  Each boundary
polytope is a portion of the boundary for which there is a single transformation matrix that maps the 
nearby non-Niggli-reduced cells to Niggli-reduced cells.  This is not necessarily a mapping back to the starting
point, not even to a point near the starting point, not even for points on the boundary polytope itself.
We look for point-by-point invariance in defining special-position subspaces
in section \ref{sec:invar}.

Each boundary polytope ${\mathitbf{\Gamma}}$ has an associated ``projector'' $P_{\mathitbf{\Gamma}}$, which is the
linear transformation that maps an arbitrary $g \in {\mathbi{G}^{\mathbi{6}}}$ to the point $P_{\mathitbf{\Gamma}} g$ closest to $g$ in the hyperplane containing ${\mathitbf{\Gamma}}$.  It is important to understand that $P_{\mathitbf{\Gamma}} g$ may not be Niggli reduced, nor even close to the Niggli cone.

\section{Special-Position Subspaces}
\label{sec:invar}

In an analysis of symmetry, special positions play an important role.
A special position is a point invariant under a symmetry transformation,
an eigenvector, of eigenvalue 1, of the transformation.  We define a
special-position subspace of a boundary polytope, ${\mathitbf{\Gamma}}$, as the 
intersection of the eigenspace of eigenvalue 1 of $M_{{\mathitbf{\Gamma}}}$ with the boundary polytope. 
Formally, for a boundary polytope, ${\mathitbf{\Gamma}}$,
with associated transformation matrix, $M_{{\mathitbf{\Gamma}}}$, the special-position subspace ${\mathitbf{\Lambda}} ({\mathitbf{\Gamma}},M_{{\mathitbf{\Gamma}}})$ is defined as
the set of points $\overrightarrow{\gamma}\ \in {\mathitbf{\Gamma}}$ such that $M_{{\mathitbf{\Gamma}}} (\overrightarrow{\gamma}) = \overrightarrow{\gamma}$.
In the case of boundary polytopes
associated with a transformation matrix that goes from the all acute $+ + +$ case to the all obtuse $- - -$ case
or vice versa, there cannot be any Niggli-reduced special-position subspace, because the axial planes of
the $g_{\{4,5,6\}}$ subspace are excluded from the all acute $+ + +$ case.    As we will see,
while the special-position subspaces of the boundary polytopes are not needed in order to classify the
primitive lattice types, they come into play in classifying the non-primitive lattice types.  

If we have two boundary polytopes ${\mathitbf{\Gamma}}_1, {\mathitbf{\Gamma}}_2 \subset  {\mathbi{N}}$, we denote the intersection of the
closure of ${\mathitbf{\Gamma}}_1$ and the closure of ${\mathitbf{\Gamma}}_2$ as ${\mathitbf{\Gamma}}_1 {\mathitbf{\Gamma}}_2$.   Note that intersection is a commutative
operation, {\it i.e.} ${\mathitbf{\Gamma}}_1 {\mathitbf{\Gamma}}_2 = {\mathitbf{\Gamma}}_2 {\mathitbf{\Gamma}}_1$.

Because we have
restricted the boundary polytope in this paper to have only one associated matrix, we use the notation
$\hat{{\mathitbf{\Gamma}}}$ for ${\mathitbf{\Lambda}} ({\mathitbf{\Gamma}},M_{{\mathitbf{\Gamma}}})$.  In general, an infinite
number of higher dimensional polytopes will intersect ${\mathitbf{\Gamma}}$ in $\hat{{\mathitbf{\Gamma}}}$.   We distinguish
such a higher dimensional polytope as ${\mathitbf{\Gamma}}^\prime$.  Thus ${\mathitbf{\Gamma}} {\mathitbf{\Gamma}}^\prime = \hat{{\mathitbf{\Gamma}}}$.

\section{The fifteen 5-D boundary polytopes}

The fifteen 5-dimensional boundary polytopes and their special-position subspaces may be organized as shown in Table \ref{5D}, 
in which we use the hexadecimal digits 1 through F as identifiers.   For each 5-dimensional  boundary polytope
${\mathitbf{\Gamma}}$ in Table \ref{5D} having a non-trivial special-position subspace, we designate the  particular choice
of higher dimensional polytope intersecting ${\mathitbf{\Gamma}}$ in $ \hat{{\mathitbf{\Gamma}}}$ as ${\mathitbf{\Gamma}}^\prime$.  See section \ref{sec:fdcase}
for a concrete example.  

In the discussions of the fifteen 5-dimensional boundary polytopes, below, we give the condition being satisfied, the right-handed
${\mathbi{E}^{\mathbi{3}}}$ presentation of the boundary transformation cell-edge-by-cell-edge, a ${\mathbi{G}^{\mathbi{6}}}$
matrix presentation of the same boundary transformation and a ${\mathbi{G}^{\mathbi{6}}}$ matrix presentation of a projector
into the hyperplane of that boundary.  Note that both a right-handed ${\mathbi{E}^{\mathbi{3}}}$ presentation and its negative (left-handed)
presentation would map into the same ${\mathbi{G}^{\mathbi{6}}}$ presentation.

\subsubsection{Equal cell-edge cases: }

Cases 1 and 2 arise when two cell edges have equal
length.   The conditions of Niggli reduction impose a secondary condition on the associated
angles for those two edges that resolves the ambiguity in ordering them. For example, in
case 1, $\|\overrightarrow{a}\| = \|\overrightarrow{b}\|$ ($g_1 = g_2$) and the Niggli-reduced  ${\mathbi{G}^{\mathbi{6}}}$ vector 
$U = (\mathbf{g_1}, \mathbf{g_2}, g_3, \mathbf{g_4}, \mathbf{g_5}, g_6)$ produces the same lattice as 
$V = (\mathbf{g_2}, \mathbf{g_1}, g_3, \mathbf{g_5}, \mathbf{g_4}, g_6)$, which is not
Niggli-reduced if $g_4$ and $g_5$ have different values. In going from Niggli-reduced
cells near $U$ to Niggli-reduced cells near $V$ (e.g. by decreasing $g_2$ slightly), we are
crossing a boundary polytope with a discontinuity in each of $g_4$ and $g_5$ \cite{Andrews1988}.  We may represent the
transformation that takes $U$ into $V$ at the first boundary polytope by the matrix $M_1$ that maps
$U$ into $V$ and the projector $P_1$ that maps any  ${\mathbi{G}^{\mathbi{6}}}$ vector into the $g_1 = g_2$ boundary polytope.
\[
\text{Case 1:~~}
g_1= g_2, ~~ \overrightarrow{a}  \rightarrow -\overrightarrow{b} , ~~ \overrightarrow{b} \rightarrow -\overrightarrow{a}, ~~ \overrightarrow{c} \rightarrow -\overrightarrow{c}
\]
\[ M_1 = \left(
{\bf 0} {\bf 1} 0 0 0 0 /
{\bf 1} {\bf 0} 0 0 0 0 /
0 0 1 0 0 0 /
0 0 0 {\bf 0} {\bf 1} 0 /
0 0 0 {\bf 1} {\bf 0} 0 /
0 0 0 0 0 1
\right)
\]
\[
P_1 =\left( 
{\bf \frac{1}{2}} {\bf \frac{1}{2}} 0 0 0 0 /
{\bf \frac{1}{2}} {\bf \frac{1}{2}} 0 0 0 0 /
0 0 1 0 0 0 /
0 0 0 1 0 0 /
0 0 0 0 1 0 /
0 0 0 0 0 1
\right)
\]
\[{\mathbi{G}^{6}}\ \text{subspace:}\ (r,r,s,t,u,v)\]
~~\\
Similarly, for case 2, $\| \overrightarrow{b} \| = \| \overrightarrow{c} \|$ ($g_2 = g_3$), $g_5$ and  $g_6$ are exchanged, yielding
\[
\text{Case 2:~~}
g_2 = g_3, ~~\overrightarrow{a}  \rightarrow -\overrightarrow{a}, ~~ \overrightarrow{b} \rightarrow -\overrightarrow{c} , ~~\overrightarrow{c}  \rightarrow -\overrightarrow{b} 
\]
\[
M_2 =\left( 
1 0 0 0 0 0  /
0 {\bf 0} {\bf 1} 0 0 0  /
0 {\bf 1} {\bf 0} 0 0 0  /
0 0 0 1 0 0  /
0 0 0 0 {\bf 0} {\bf 1}  /
0 0 0 0 {\bf 1} {\bf 0}\right)
\]
\[
P_2 =\left( 
1 0 0 0 0 0  /
0 {\bf \frac{1}{2}} {\bf \frac{1}{2}} 0 0 0  /
0 {\bf \frac{1}{2}} {\bf \frac{1}{2}} 0 0 0  /
0 0 0 1 0 0  /
0 0 0 0 1 0  /
0 0 0 0 0 1\right)\]
\[{\mathbi{G}^{6}}\ \text{subspace:}\ (r,s,s,t,u,v)\]
~~\\
The remaining cell-edge case in which $\|\overrightarrow{a}\| = \|\overrightarrow{c}\|$ ($g_1 = g_3$), is only considered
under Niggli reduction when $\|\overrightarrow{a}\| = \|\overrightarrow{b}\|$ and $\|\overrightarrow{b}\| = \|\overrightarrow{c}\|$ which is a combination of case 1
and case 2.  This requires two simultaneous 5-dimensional constraints, thereby making $g_1 = g_3$
 a 4-dimensional rather than a 5-dimensional case.
 
 The special-position subspaces $\hat{1}$ and $\hat{2}$ are obtained by adding the constraints $1^\prime\!: \{g_4 = g_5\}$ and
 $2^\prime\!:\{g_5=g_6\}$, respectively.
 
\subsubsection{Ninety-degree cases:}

 Cases 3, 4 and 5 arise when a reduced-cell angle is ninety
degrees. In those cases, the remaining cell angles both can be replaced by
their supplements.  This changes the sign of $g_{\{ 4, 5, 6 \}}$.

\[
\text{Case 3:~~}
g_4 = 0, ~~ \overrightarrow{a}  \rightarrow  \overrightarrow{a}, ~~ \overrightarrow{b}  \rightarrow -\overrightarrow{b}, ~~ \overrightarrow{c}  \rightarrow -\overrightarrow{c}
\]
\[
M_3 =\left( 
1 0 0 0 0 0  /
0 1 0 0 0 0  /
0 0 1 0 0 0  /
0 0 0 1 0 0  /
0 0 0 0 {\bf \overline{1}
} 0  /
0 0 0 0 0 {\bf \overline{1}
}\right)
\]
\[
P_3 =\left( 
1 0 0 0 0 0  /
0 1 0 0 0 0  /
0 0 1 0 0 0  /
0 0 0 {\bf 0} 0 0  /
0 0 0 0 1 0  /
0 0 0 0 0 1\right)
\]
\[{\mathbi{G}^{6}}\ \text{subspace:}\ (r,s,t,0,-u,-v)\]

\[
\text{Case 4:~~}
g_5 = 0, ~~ \overrightarrow{a}  \rightarrow  -\overrightarrow{a}, ~~ \overrightarrow{b}  \rightarrow \overrightarrow{b}, ~~ \overrightarrow{c}  \rightarrow -\overrightarrow{c}
\]
\[
M_4 =\left(
1 0 0 0 0 0  /
0 1 0 0 0 0  /
0 0 1 0 0 0  /
0 0 0 {\bf \overline{1}
} 0 0  /
0 0 0 0 1 0  /
0 0 0 0 0 {\bf\overline{1}
}\right)
\]
\[
P_4 =\left( 
1 0 0 0 0 0  /
0 1 0 0 0 0  /
0 0 1 0 0 0  /
0 0 0 1 0 0  /
0 0 0 0 {\bf 0} 0  /
0 0 0 0 0 1\right)
\]
\[{\mathbi{G}^{6}}\ \text{subspace:}\ (r,s,t,-u,0,-v)\]

\[
\text{Case 5:~~}
g_6 = 0, ~~ \overrightarrow{a}  \rightarrow  -\overrightarrow{a}, ~~ \overrightarrow{b}  \rightarrow -\overrightarrow{b}, ~~ \overrightarrow{c}  \rightarrow \overrightarrow{c}
\]
\[
M_5 =\left(
1 0 0 0 0 0  /
0 1 0 0 0 0  /
0 0 1 0 0 0  /
0 0 0 {\bf \overline{1}
} 0 0  /
0 0 0 0 {\bf \overline{1}
} 0  /
0 0 0 0 0 1\right)
\]
\[
P_5 =\left( 
1 0 0 0 0 0  /
0 1 0 0 0 0  /
0 0 1 0 0 0  /
0 0 0 1 0 0  /
0 0 0 0 1 0  /
0 0 0 0 0 {\bf 0}\right)
\]
\[{\mathbi{G}^{6}}\ \text{subspace:}\ (r,s,t,-u,-v,0)\]

In each ninety-degree case, the special-position subspace consists of $\hat{3}, \hat{4}, \hat{5}\!: \{g_4 = g_5 = g_6 = 0\}$,
{\it i.e.} the primitive orthorhombic case, and we take
$3^\prime\!: \{g_5 = g_6 = 0\}$, $4^\prime\!: \{g_4 = g_6 = 0\}$, $5^\prime\!: \{g_4 = g_5 = 0\}$.

\subsubsection{Face-diagonal cases:}
\label{sec:fdcase}

Cases 6 through E are all face-diagonal cases, in which
a cell edge is equal in length to a face-diagonal. Some complexity arises in the
analysis because, unlike Delaunay reduction, Niggli reduction permits non-obtuse
angles. We can always change the sign of any two elements of $g_{\{4,5,6\}}$ by changing the
direction of the cell edge involved with those two elements. 
For example, if we transform $ \overrightarrow{a}$  to $ - \overrightarrow{a}$ then, then while $g_1$ remains
unaffected, the signs of each of $g_5 = 2 \overrightarrow{a}\!\!\cdot\!\overrightarrow{c}$ and $g_6 = 2 \overrightarrow{a}\!\!\cdot\!\overrightarrow{b}$ will change.
Thus we can transform a cell having three acute angles to a cell having 
one acute angle and two obtuse angles, and we can transform a cell
having three obtuse angles to a cell having one obtuse angle and two acute
angles.   The complete list of possible sign changes in  $g_{\{4,5,6\}}$ by changing the directions
of axes are:
\[
 \{+ + +\}  ~~\leftrightarrows~~ \{- - +\} ~~\leftrightarrows~~ \{- + -\} ~~\leftrightarrows~~ \{+ - -\}
 \]
 \[
 \{- - -\} ~~\leftrightarrows~~ \{+ + -\} ~~\leftrightarrows~~ \{+ - +\} ~~\leftrightarrows~~ \{- + +\}
\]

Unless one of the angles is ninety degrees (which introduces a zero into $g_{\{4,5,6\}}$),
we cannot ordinarily transform a Niggli-reduced cell with all-acute angles to one with 
all-obtuse angles by changing the directions of axes, nor can we transform a Niggli-reduced cell with 
all-obtuse angles to one with all-acute angles by changing the directions of the axes.  
Note that changing the direction of all three axes has no effect because all the sign changes cancel.

The face-diagonal cases do include cases in which transformations from, for example, \(+ + +\)
to \(- - -\) do occur.   Let us look in detail at cases 6 and 7, 
$ g_2 = g_4$. 
\begin{eqnarray*}
 g_2  &=& g_4 \\
 g_2 - g_4 &=& 0\\
 g_2  - g_4 + g_3 &=& g_3\\
\overrightarrow{b}\!\!\cdot\!\overrightarrow{b} - 2  \overrightarrow{b}\!\!\cdot\!\overrightarrow{c}  + \overrightarrow{c}\!\!\cdot\!\overrightarrow{c}
 &=& \overrightarrow{c}\!\!\cdot\!\overrightarrow{c}\\
 \|  \overrightarrow{b}  -  \overrightarrow{c} \|^2 &=& \| \overrightarrow{c} \|^2\\
 \|  \overrightarrow{b}  -  \overrightarrow{c} \| &=& \| \overrightarrow{c} \|\\
\end{eqnarray*}
Thus transforming $\overrightarrow{c}$ to $\overrightarrow{b} - \overrightarrow{c}$ will not change the cell edge lengths. In this case,
$g_1$, $g_2$, and $g_3$ are, of course, unchanged and
\begin{eqnarray*}
g_4' &=& 2 \overrightarrow{b}\!\!\cdot\!( \overrightarrow{b} - \overrightarrow{c} ) = 2 \overrightarrow{b}\!\!\cdot\!  \overrightarrow{b} - 2 \overrightarrow{b}\!\!\cdot\!\overrightarrow{c} = 2 g_2 - g_4 = g_4\\
g_5' &=& 2 \overrightarrow{a} \!\!\cdot\!( \overrightarrow{b} - \overrightarrow{c}) = 2 \overrightarrow{a}\!\!\cdot\!  \overrightarrow{b}  -  2 \overrightarrow{a} \!\!\cdot\!\overrightarrow{c} =  g_6  -  g_5\\
g_6' &=& 2 \overrightarrow{a}\!\!\cdot\! \overrightarrow{b} =  g_6
\end{eqnarray*}
This shows that a single element of $g_{\{4,5,6\}}$, $g_5$, will change sign depending on the sign of  $g_6 - g_5$. 
Cases 6 and 7 cannot start from the all-obtuse case because $g_4 = g_2$ and because $g_2$ must be nonnegative.
Starting from an all acute case, \(+ + +\), we will remain in the all acute case if  $g_6$ is greater
than or equal to $g_5$ but go to one having one obtuse angle (not reduced) if  $g_6$ is less than $g_5$. We
then change to having all obtuse angles, \(- - -\), by reversing the direction of $b$.
The resulting matrices in these face-diagonal cases are:
\[
\text{Case 6:~~}
g_2 = g_4, ~~ g_5 \geqslant  g_6, ~~\overrightarrow{a}  \rightarrow \overrightarrow{a}, ~~ \overrightarrow{b}  \rightarrow - \overrightarrow{b} , ~~\overrightarrow{c}   \rightarrow \overrightarrow{b} - \overrightarrow{c}
\]
\[
M_6 =\left(
1 0 0 0 0 0  /
0 1 0 0 0 0  /
0 1 1 \overline{1}
 0 0  /
0 \overline{2} 0 1 0 0  /
0 0 0 0 \overline{1}
 1  /
0 0 0 0 0 \overline{1}
\right)
\]
\[
P_6 =\left(
1 0 0 0 0 0  /
0 {\bf \frac{1}{2}} 0 {\bf \frac{1}{2}} 0 0  /
0 0 1 0 0 0  /
0 {\bf \frac{1}{2}} 0 {\bf \frac{1}{2}} 0 0  /
0 0 0 0 1 0  /
0 0 0 0 0 1  \right)
\]
\[{\mathbi{G}^{6}}\ \text{subspace:}\ (r,s,t,s,u+v,v)\]

\[
\text{Case 7:~~}
g_2 = g_4, ~~g_5  <   g_6, ~~\overrightarrow{a}  \rightarrow - \overrightarrow{a}, ~~\overrightarrow{b}  \rightarrow - \overrightarrow{b}, ~~\overrightarrow{c}  \rightarrow \overrightarrow{c} - \overrightarrow{b}
\]
\[
M_7 =\left(
1 0 0 0 0 0  /
0 1 0 0 0 0  /
0 1 1 \overline{1}
 0 0  /
0 2 0 \overline{1}
 0 0  /
0 0 0 0 \overline{1}
 1  /
0 0 0 0 0 1\right)
\]
\[
P_7 =\left(
1 0 0 0 0 0  /
0 {\bf \frac{1}{2}} 0 {\bf \frac{1}{2}} 0 0  /
0 0 1 0 0 0  /
0 {\bf \frac{1}{2}} 0 {\bf \frac{1}{2}} 0 0  /
0 0 0 0 1 0  /
0 0 0 0 0 1  \right)
\]
\[{\mathbi{G}^{6}}\ \text{subspace:}\ (r,s,t,s,u,u+v)\]

\[
\text{Case 8:~~}
g_2 = -g_4, ~~\overrightarrow{a}  \rightarrow \overrightarrow{a}, ~~\overrightarrow{b}  \rightarrow -\overrightarrow{b}, ~~\overrightarrow{c} \rightarrow -\overrightarrow{b} -\overrightarrow{c}
\]
\[
M_8 =\left(
1 0 0 0 0 0  /
0 1 0 0 0 0  /
0 1 1 1 0 0  /
0 2 0 1 0 0  /
0 0 0 0 \overline{1}
 \overline{1}
  /
0 0 0 0 0 \overline{1}
\right)
\]
\[
P_8 =\left(
1 0 0 0 0 0  /
0 {\bf \frac{1}{2}} 0 {\bf \frac{\overline{1}}{2}} 0 0  /
0 0 1 0 0 0  /
0 {\bf \frac{\overline{1}}{2}} 0 {\bf \frac{1}{2}} 0 0  /
0 0 0 0 1 0  /
0 0 0 0 0 1  \right)
\]
\[{\mathbi{G}^{6}}\ \text{subspace:}\ (r,s,t,-s,-u,-v)\]

\[
\text{Case 9:~~}
g_1= g_5, ~~g_4 \geqslant  g_6, ~~\overrightarrow{a}  \rightarrow -\overrightarrow{a}, ~~\overrightarrow{b}  \rightarrow  \overrightarrow{b}, ~~\overrightarrow{c}   \rightarrow \overrightarrow{a} - \overrightarrow{c}
\]
\[
M_9 =\left(
1 0 0 0 0 0 /
0 1 0 0 0 0  /
1 0 1 0 \overline{1}
 0  /
0 0 0 \overline{1}
 0 1  /
\overline{2} 0 0 0 1 0  /
0 0 0 0 0 \overline{1}
\right)
\]
\[
P_9 =\left(
{\bf \frac{1}{2}} 0 0 0 {\bf \frac{1}{2}} 0  /
0 1 0 0 0 0  /
0 0 1 0 0 0  /
0 0 0 1 0 0  /
{\bf \frac{1}{2}} 0 0 0 {\bf \frac{1}{2}} 0  /
0 0 0 0 0 1\right)
\]
\[{\mathbi{G}^{6}}\ \text{subspace:}\ (r,s,t,u+v,r,u)\]

\[
\text{Case A:~~}
g_1= g_5, ~~g_4  <   g_6, ~~\overrightarrow{a}  \rightarrow  -\overrightarrow{a}, ~~\overrightarrow{b}  \rightarrow -\overrightarrow{b}, ~~\overrightarrow{c}  \rightarrow -\overrightarrow{a} + \overrightarrow{c}
\]
\[
M_{A} =\left(
1 0 0 0 0 0  /
0 1 0 0 0 0  /
1 0 1 0 \overline{1}
 0  /
0 0 0 \overline{1}
 0 1  /
2 0 0 0 \overline{1}
 0  /
0 0 0 0 0 1\right)
\]
\[
P_{A} =\left(
{\bf \frac{1}{2}} 0 0 0 {\bf \frac{1}{2}} 0  /
0 1 0 0 0 0  /
0 0 1 0 0 0  /
0 0 0 1 0 0  /
{\bf \frac{1}{2}} 0 0 0 {\bf \frac{1}{2}} 0  /
0 0 0 0 0 1\right)
\]
\[{\mathbi{G}^{6}}\ \text{subspace:}\ (r,s,t,u,r,u+v)\]

\[
\text{Case B:~~}
g_1= -g_5, ~~\overrightarrow{a}  \rightarrow  -\overrightarrow{a}, ~~\overrightarrow{b}  \rightarrow \overrightarrow{b} , ~~\overrightarrow{c}   \rightarrow -\overrightarrow{a} - \overrightarrow{c}
\]
\[
M_{B} =\left(
1 0 0 0 0 0  /
0 1 0 0 0 0  /
1 0 1 0 1 0  /
0 0 0 \overline{1}
 0 \overline{1}  /
2 0 0 0 1 0  /
0 0 0 0 0 \overline{1}\right)
\]
\[
P_{B} =\left(
{\bf \frac{1}{2}} 0 0 0 {\bf \frac{\overline{1}}{2}} 0  /
0 1 0 0 0 0  /
0 0 1 0 0 0  /
0 0 0 1 0 0  /
{\bf \frac{\overline{1}}{2}} 0 0 0 {\bf \frac{1}{2}} 0  /
0 0 0 0 0 1\right)
\]
\[{\mathbi{G}^{6}}\ \text{subspace:}\ (r,s,t,-u,-r,-v)\]

\[
\text{Case C:~~}
g_1 =  g_6,g_4 \geqslant g_5, ~~\overrightarrow{a}  \rightarrow  -\overrightarrow{a} , ~~\overrightarrow{b}   \rightarrow \overrightarrow{a} - \overrightarrow{b}, ~~\overrightarrow{c}  \rightarrow  \overrightarrow{c}
\]
\[
M_{C} =\left(
1 0 0 0 0 0  /
1 1 0 0 0 \overline{1}  /
0 0 1 0 0 0  /
0 0 0 \overline{1} 1 0  /
0 0 0 0 \overline{1} 0  /
\overline{2} 0 0 0 0 1\right)
\]
\[
P_{C} =\left( 
{\bf \frac{1}{2}} 0 0 0 0 {\bf \frac{1}{2}}  /
0 1 0 0 0 0  /
0 0 1 0 0 0  /
0 0 0 1 0 0  /
0 0 0 0 1 0  /
{\bf \frac{1}{2}} 0 0 0 0 {\bf \frac{1}{2}}\right)
\]
\[{\mathbi{G}^{6}}\ \text{subspace:}\ (r,s,t,u+v,v,r)\]

\[
\text{Case D:~~} g_1 =  g_6, ~~g_4 < g_5, ~~\overrightarrow{a}  \rightarrow -\overrightarrow{a}, ~~ \overrightarrow{b}   \rightarrow -\overrightarrow{a} + \overrightarrow{b}, ~~\overrightarrow{c}  \rightarrow - \overrightarrow{c}
\]
\[
M_{D} = \left( 
1 0 0 0 0 0  /
1 1 0 0 0 \overline{1}  /
0 0 1 0 0 0  /
0 0 0 \overline{1} 1 0  /
0 0 0 0 1 0  /
2 0 0 0 0 \overline{1}\right)
\]
\[
P_{D} =\left( 
{\bf \frac{1}{2}} 0 0 0 0 {\bf \frac{1}{2}}  /
0 1 0 0 0 0  /
0 0 1 0 0 0  /
0 0 0 1 0 0  /
0 0 0 0 1 0  /
{\bf \frac{1}{2}} 0 0 0 0 {\bf \frac{1}{2}}\right)
\]
\[{\mathbi{G}^{6}}\ \text{subspace:}\ (r,s,t,u,u+v,r)\]

\[
\text{Case E:~~} g_1= - g_6, ~~\overrightarrow{a}  \rightarrow -\overrightarrow{a}, ~~  \overrightarrow{b}   \rightarrow -\overrightarrow{a} - \overrightarrow{b} , ~~ \overrightarrow{c}  \rightarrow \overrightarrow{c}
\]
\[
M_{E} = \left( 
1 0 0 0 0 0  /
1 1 0 0 0 1  /
0 0 1 0 0 0  /
0 0 0 \overline{1} \overline{1} 0  /
0 0 0 0 \overline{1} 0  /
2 0 0 0 0 1\right)
\]
\[
P_{E} =\left( 
{\bf \frac{1}{2}} 0 0 0 0 {\bf \frac{\overline{1}}{2}}  /
0 1 0 0 0 0  /
0 0 1 0 0 0  /
0 0 0 1 0 0  /
0 0 0 0 1 0  /
{\bf \frac{\overline{1}}{2}} 0 0 0 0 {\bf \frac{1}{2}}\right)
\]
\[{\mathbi{G}^{6}}\ \text{subspace:}\ (r,s,t,-u,-v,-r)\]

The special-position subspaces of the face-diagonal boundary polytopes 6, 8, 9, B, C and E are empty because such a
special position would require a common point in the all acute $+ + +$ and all obtuse $- - -$ cases that only meet
at the axial planes of the $g_{\{4,5,6\}}$ subspace, which are excluded from the all acute $+ + +$ case.  For cases
7, A and D there are non-trivial special-position subspaces. 
An invariant point in case 7 would have to satisfy $g_5 = g_6 - g_5$ or $g_5 = g_6/2$.
Thus we define $7^\prime\!: \{g_5 = g_6/2\}$ and similarly define $A^\prime\!: \{g_4 = g_6/2\}$ and $D^\prime\!: \{g_4 = g_5/2\}$.

\subsubsection{Body-diagonal case:}

There is only one 5-dimensional body-diagonal case, 
$\| \overrightarrow{a} + \overrightarrow{b} + \overrightarrow{c} \| = \| \overrightarrow{c} \|:$
\[
\text{Case F:~~}  g_1+ g_2 + g_3  + g_4 + g_5 +  g_6 = g_3, ~~\overrightarrow{a}  \rightarrow - \overrightarrow{a} ,~~\overrightarrow{b}  \rightarrow - \overrightarrow{b} ,~~\overrightarrow{c}  \rightarrow \overrightarrow{a} +\overrightarrow{b} + \overrightarrow{c}
\]
\[
M_{F} =\left( 
1 0 0 0 0 0  /
0 1 0 0 0 0  /
1 1 1 1 1 1  /
0 \overline{2} 0 \overline{1} 0 \overline{1}  /
\overline{2} 0 0 0 \overline{1} \overline{1}  /
0 0 0 0 0 1\right)
\]
\[
P_{F} =\left( 
\frac{4}{5} \frac{\overline{1}}{5} 0 \frac{\overline{1}}{5} \frac{\overline{1}}{5} \frac{\overline{1}}{5}  /
\frac{\overline{1}}{5} \frac{4}{5} 0 \frac{\overline{1}}{5} \frac{\overline{1}}{5} \frac{\overline{1}}{5}  /
0 0 1 0 0 0  /
\frac{\overline{1}}{5} \frac{\overline{1}}{5} 0 \frac{4}{5} \frac{\overline{1}}{5} \frac{\overline{1}}{5}  /
\frac{\overline{1}}{5} \frac{\overline{1}}{5} 0 \frac{\overline{1}}{5} \frac{4}{5} \frac{\overline{1}}{5}  /
\frac{\overline{1}}{5} \frac{\overline{1}}{5} 0 \frac{\overline{1}}{5} \frac{\overline{1}}{5} \frac{4}{5}  /
\right)
\]
\[{\mathbi{G}^{6}}\ \text{subspace:}\ (r,s,t,-u,-v,-r-s+u+v)\]

In order to have a special-position subspace in case F, in addition to $g_1+ g_2 + g_3  + g_4 + g_5 +  g_6 = g_3$, we need (from the
fourth and fifth rows of $M_F$), $g_4 = -2 g_2 - g_4 -g_6$ and $g_5 = -2 g_1 - g_5 - g_6$ from which we have 
$2 g_2 + 2 g_4 = -g_6 = 2 g_1 + 2 g_5$ from which we take $F^\prime\!:\{g_1 - g_2 - g_4+g_5 = 0\}$.  
This is equivalent to
\(\|\overrightarrow{a}+\overrightarrow{c}\| = \|\overrightarrow{b}+\overrightarrow{c}\|\), \textit{i.e.}
that the shorter b-face-diagonal is the same length as the shorter  a-face-diagonal.

\section{The 4-D boundary polytopes}

The 4-D boundary polytopes are created by the intersection of 2 5-D boundary polytopes.  
Certain intersections are degenerate.  For example, cases  8, B, E and F are restricted
to the $- - -$ branch of the boundary of the Niggli cone while cases 6, 7, 9, A, C and D are restricted to the $+ + +$ branch.
More subtly, cases 6 and 7
require $g_2 = g_4$, maximizing $g_4$,  Forcing $g_4 \geqslant g_5$ and $g_4 \geqslant g_6$ 
which would conflict with cases A and D.  Only 
cases 9 and C, respectively, taken to their boundaries with A and D, respectively are possible.
Those boundaries are designated 9A and CD, respectively.  
Similarly, cases 9 and A maximize $g_5$, which would conflict with case 7 and force 6 to the 67 boundary, and cases 
C and D maximize $g_6$ which would conflict with case 6 except at the 67 boundary.   Thus cases
6A, 7A, 6D, 7D, 79, 7A, 6C, 6D actually are the lower dimension cases 69A, 79A, 6CD, 7CD,
79A, 79A, 6CD and 6CD, respectively.  This process can result in 3-D or even 2-D boundary polytopes
from the intersection of two 5-D boundary polytopes (see below).    After
excluding the cases that involve any of $g_{\{1,2,3\}} = 0$, there are fifty-five 4-D cases
as shown in Appendix \ref{sec:mc} Table \ref{4D}.  The relative populations for all the 2-D boundary polytopes except
26,
28,
2A and 
2D
have Z-scores above -1.  The Z-scores for those 4 cases range from -1.1 down to -1.9. (See the supplementary
materials \ref{sec:mc} for a discussion of Z-scores).

The edges of the 5-D polytopes can be read directly from Appendix \ref{sec:mc} Table \ref{4D}.  For example the
6 polytope is bounded by 16, 26, 56, 67 and 69 and the F polytope is bounded by 1F, 2F, 8F,
BF and EF.   It is important to note that the polytopes 1, 2, 3, 4 and 5  extend into the boundaries
of both the $+ + +$ and $- - -$ branches of ${\mathbi{N}}$.  Even though the polytopes 3, 4 and 5 do not
contain any valid Niggli reduced $+ + +$ cells, they are part of both branches of $\partial({\mathbi{N}})$ even for $+ + +$.

\section{The 3-D boundary polytopes}

The 3-D boundary polytopes are created by the intersection of 3 5-D boundary polytopes.  In some cases the
boundary polytope is better represented by a 4-fold intersection.  The boundary polytope 34CD is equivalent to 34C and 34D,
359A is equivalent to 359 and 35A, 4567 is equivalent to 456 and 457,
679C is equivalent to 69C and 79C, and
9ACD is equivalent to 9AD and ACD.  
These are ``flat boundary intersection'' cases in
which one side of the flat boundary intersection implies the other.  On the other hand
126 and 127, 12A and 129, 12C and 12D, 2AD and 29C, 69C and 79C 
are distinct rather than equivalent pairs of flat boundary intersections.
Six 2-fold intersections of 5-D boundary polytopes (6A, 6C, 79, 7D, 9D and AC) result in
3-D boundary polytopes, rather than in 4-D boundary polytopes.   In each case both boundary polytopes
have mismatched partners from ``flat boundary intersections''.   Let us examine the 6A case in detail, elaborating
on the discussion of the 4-D boundary polytopes above.

Cases 6 and A  are $g_2 = g_4$ and $g_5 \geqslant  g_6$ and $g_1 = g_5 $ and $g_4 < g_6$,
respectively.  For intersections, we use the closures of the boundary polytopes, so
we have the closure of A as $g_1 = g_5 $ and $g_4 \leqslant g_6$.
The Niggli cone itself imposes the additional restrictions $g_6 \leqslant g_1$ and $g_1 \leqslant g_2$,
but from 6, $g_2 = g_4$ and from the closure of A, $g_4 \leqslant g_6$, so we have
\[ g_2 = g_4 \leqslant g_6 \leqslant g_1 \Rightarrow \ g_2 \leqslant g_1 \]
and from the Niggli reduction conditions
\[ g_1 \leqslant  g_2 \]
thus $g_1 = g_2$ and $g_4 = g_6$, meaning that, in addition to satisfying the constraints
of case A, we also satisfy the constraints of case 9, $g_1 = g_5 $ and $ g_4 \geqslant  g_6$.
Thus case 6A is actually case 69A, producing a true 3-D boundary polytope from the intersection
of 2 5-D boundary polytopes due to the additional constraints of Niggli reduction.  As we will see
in the discussion of the 2-D boundary polytopes, the Niggli reduction constraints can result in shedding
one or more degrees of freedom, allowing some 2-fold intersections of 5-D boundary polytopes to
result in 2-D boundary polytopes.

 After
excluding the cases that involve any of $g_{\{1,2,3\}} = 0$, there are seventy-nine cases as shown
in Appendix \ref{sec:mc} Table \ref{3D}.  The relative populations for all the 3-D boundary polytopes have
Z-scores greater than -1.1. 

For completeness, if one wished to include the boundary polytope with $g_1 = 0$
it would be considered in the 3-D polytopes.  In $cl({\mathbi{N}})$, $g_1=0$ forces $g_5=g_6=0$, $g_2=0$,
leaving only 3 degrees of freedom at $\partial({\mathbi{N}})$.  Note that $g_2=0$ is of even lower dimension,
1, because $g_2 = 0$ forces $g_4=0$ as well as $g_1=g_5=g_6=0$, leaving only 1 degree of
freedom ($g_3$).  $g_3=0$ is just the origin.

\section{The 2-D boundary polytopes}

The 2-D boundary polytopes are, in general, created by the intersection of  4
5-D boundary polytopes, but several well-populated 4-fold intersections
result in 3-D boundary polytopes rather than 2-D boundary polytopes.  Several 4-fold intersections 
are most naturally presented as higher multiplicity intersections, and in some cases
the intersection of as few as 2 5-D boundary polytopes is sufficient to create a
2-D boundary polytope.     After excluding the cases that involve
any of $g_{\{1,2,3\}} = 0$, there are fifty-five cases
as shown Appendix \ref{sec:mc} Table \ref{2D}. 
The combination of 7 boundary polytopes 1679ACD 
is the hexagonal rhombohedral $hR$, lattice character 9,
Roof/Niggli symbol 49B, subspace $( r, r, s, r, r, r)$.  Alternatively, lattice character 9 can be viewed as any of 
81 other intersections, including 2 2-folds (6D, 7A), 18 3-folds (179,
16A,
16C,
16D,
17A,
17D,
19D,
1AC,
67A,
67D,
69D,
6AC,
6AD,
6CD,
79A,
79D,
7AC,
7AD),  33 4-folds (1679
167A,
167C,
167D,
169A,
169C,
169D,
16AC,
16AD,
16CD,
179A,
179C,
179D,
17AC,
17AD,
17CD,
19AC,
19AD,
19CD,
1ACD,
679A,
679D,
67AC,
67AD,
67CD,
69AC,
69AD,
69CD,
6ACD,
79AC,
79AD,
79CD,
7ACD), 21 5-folds
(1679A,
1679C,
1679D,
167AC,
167AD,
167CD,
169AC,
169AD,
169CD,
16ACD,
179AC,
179AD,
179CD,
17ACD,
19ACD,
679AC,
679AD,
679CD,
67ACD,
69ACD,
79ACD) and 7 6-folds (1679AC,
1679AD,
1679CD,
167ACD,
169ACD,
179ACD,
679ACD)
 and is a very highly populated
2-D boundary polytope.  If we exclude this case, the  remaining fifty-four cases have
Z-scores ranging from -1.36 (for 1456) to 0.84 (for 123E) to 2.75 (for 29ACD).

Let us consider how one of the 2-folds, 6D, results in only 2-degrees of freedom.
Cases 6 and D are $g_2 = g_4 $ and $ g_5 \geqslant  g_6$ and 
$g_1 =  g_6 $ and $ g_4 < g_5$, respectively.  The closure of D is
$g_1 =  g_6 $ and $ g_4 \leqslant g_5$,  and Niggli reduction requires
$g_6 \leqslant g_1 \leqslant g_2$, from which it follows that
\[g_6  = g_1 \leqslant g_2 = g_4  \leqslant g_5  \leqslant g_1\]
and therefore
\[ g_6 = g_1 = g_2 = g_4 = g_5\]
creating the subspace $(r, r, s, r, r, r)$, {\it i.e.} 2 degrees of freedom.

\section{The 1-D boundary polytopes}

There are 14 distinct 1-D boundary polytopes, with many equivalent presentations.
The most complex situation is best presented as an 8-fold intersection, 12679ACD,
{\it i.e.} $g_1 = g_2 = g_3 = g_4 = g_5 =g_6$, which is the face-centered cubic $(r,r,r,r,r,r)$.
There are 81 other equivalent presentations of the face-centered cubic, inherited from the 7-fold intersection
hexagonal rhombohedral discussed above by adding case 2 to each of those presentations, thereby providing
2 3-folds for the face centered cubic (26D and 27A).
The remaining 13 1-D boundary polytopes are
12345 $(r,r,r,0,0,0)$,
1234CD $(r,r,r,0,0,r)$ (equivalent to 1234C, 1234D, 123CD, 124CD),
1234E $(r,r,r,0,0,-r)$,
12359A $(r,r,r,0,r,0)$ (equivalent to 12359, 1235A, 1239A, 1259A),
1235B $(r,r,r,0,-r,0)$,
123AD $(r,r,r,0,r,r)$,
123BEF $(r,r,r,0,-r,-r)$ (equivalent to 123BE,  123BF, 123EF, 12BEF, 23BEF),
124567 $(r,r,r,r,0,0)$ (equivalent to 12456, 12457, 12467, 12567),
12458 $(r,r,r,-r,0,0)$,
1247C $(r,r,r,r,0,r)$,
1248EF $(r,r,r,-r,0,-r)$ (equivalent to 1248E,  1248F, 124EF, 128EF),
12569 $(r,r,r,r,r,0)$,
1258BF $(r,r,r,-r,-r,0)$ (equivalent to 1258B, 1258F, 125BF, 128BF).

These 14 1-D boundary polytopes of ${\mathbi{N}}$ correspond exactly to the 14 vertices of the hyperpolyhedra
in \cite[Table 1]{Gruber1997} for which none of $g_{\{1,2,3\}}$ are zero.   The 14 that match are a
confirmation of the completeness of this analysis. Due to the distortion introduced by projection, the rejection of
the cases for which any of 
$g_{\{1,2,3\}}$ are zero is important in preserving the metric for incommensurate edges near the origin.

If we exclude the highly populated face-centered cubic, the
Z-scores for the relative populations of the remaining  13 1-D boundary polytopes range from 
more than -1.1 for 12569 to 1.44 for 123BEF.

\section{Relationship between boundary polytopes and lattice-type}
\label{sec:rel}

``\textbf{Lattice characters}'' provide a finer-grained division of lattice type than the 14 Bravais lattice types
 \cite[International Tables, Volume A]{Burzlaff1992}.
In order to understand the relationship among between the 216 ${\mathbi{G}^{\mathbi{6}}}$ boundary polytopes and
the 44 lattice characters in the International Tables,
we use combinations of the 15 5-D boundary polytopes and of the special-position subspaces
of those polytopes.  There are multiple alternate representations of some of the
lattice characters.  We discuss some of them below.
 
We refer to Roof's redrawn Niggli figure identifiers \cite{Roof1967} as ``\textbf{Roof/Niggli symbols}." 
We may associate the Roof/Niggli symbols, lattice characters and Bravais lattice types with the indicated
subspaces of ${\mathbi{G}^{\mathbi{6}}}$ and combinations of boundary polytopes and special conditions as 
shown in Tables \ref{NiggliFormsI} and 
\ref{NiggliFormsII}.  The triclinic lattice characters 31 and 44 are not included because 
no boundary polytopes are needed for the triclinic case as they fill the Niggli-reduced cone.

The primitive Bravais lattice types have a simple relationship to the boundary polytopes.  The primitive
cubic, which has one degree of freedom as a ${\mathbi{G}^{\mathbi{6}}}$ subspace, is the intersection of five 5-D boundary polytopes.
The primitive tetragonal and primitive hexagonal lattice types each have two degrees of freedom as
${\mathbi{G}^{\mathbi{6}}}$ subspaces and each is the intersection of four 5-D boundary polytopes.  The primitive orthorhombic
lattice type has three degrees of freedom as a ${\mathbi{G}^{\mathbi{6}}}$ subspace and is the intersection of three
5-D boundary polytopes.  The primitive monoclinic lattice types each have four degrees of freedom as
${\mathbi{G}^{\mathbi{6}}}$ subspaces, and each is the intersection of two 5-D boundary polytopes.

The combination of 8 boundary polytopes 12679ACD
(equivalent to the 3-fold combinations
26D and 27A) is the face-centered cubic $cF$, lattice character 1,
Roof/Niggli symbol 44C,
subspace $( r, r, r, r, r, r)$ \cite{Andrews1988}.  Alternatively, $cF$ can be viewed as any of $\hat{1}\hat{2}7$,
$\hat{1}\hat{2}A$ or $\hat{1}\hat{2}D$ and of several other intersections.
As one would expect from the large number of intersecting boundary polytopes, this is a very complex region
of ${\mathbi{G}^{\mathbi{6}}}$ and will be the subject of a later paper.

The 1-D combinations of 6 boundary polytopes 1234CD (equivalent to the 5-folds 1234C,
1234D,
123CD,
124CD), 12359A
(equivalent to the 5-folds
12359,
1235A,
1239A,
1259A),
123BEF
(equivalent to the 5-folds
123BE,
123BF,
123EF,
12BEF,
23BEF), 
124567
(equivalent to the 5-folds
12456,
12457,
12467,
12567),
1248EF
(equivalent to the 5-folds
1248E,
1248F,
124EF,
128EF), 
1258BF
(equivalent to the 5-folds
1258B,
1258F,
125BF,
128BF) form the
subspaces
\[
\begin{array}{l}
  (r, r, r, 0, 0, r)\\
  (r, r, r, 0, r, 0)\\
  (r, r, r, 0, -r, -r)\\
  (r, r, r, r, 0, 0)\\
  (r, r, r,-r, 0, -r)\\
  (r, r, r,-r, -r, 0)\\
  \end{array}
\]
of which none are Niggli-reduced and are therefore a set of open boundary polytopes.

Most of the 216 ${\mathbi{G}^{\mathbi{6}}}$ boundary polytopes are non-Niggli-reduced open boundary polytopes of the Niggli region.
Therefore only two of the 5-fold boundary polytopes,  five of the 4-fold boundary polytopes,
eight of the 3-fold boundary polytopes and eleven of the  2-fold boundary polytopes
correspond directly to lattice characters.  None of the single 5-D boundary polytopes  corresponds
to lattice characters.

We are working from the boundary polytopes, looking for the resulting symmetries.  Hosoya \cite{Hosoya1990}
started, instead, from the three highest symmetry lattice types (the three cubics) and
added the lower symmetry primitive hexagonal lattice type, and inferred boundary polytopes from the symmetries,
having to treat open boundary polytopes as if they were Niggli-reduced.    The resulting six boundary polytopes
of the three cubics and the primitive hexagonal restated in terms of ${\mathbi{G}^{\mathbi{6}}}$ are given in Table
\ref{hosoyaboundaries}.

\section{Summary and Conclusions}

The wide-spread use of Niggli reduction in crystallography implies that it
should be thoroughly understood.  Robust identification of Bravais
lattices and lookup of unit cell parameters in databases would be improved
if a successful embedding of the Niggli space could be achieved. We have
investigated and enumerated the several kinds of boundary polytopes on the Niggli
cone and enumerated the transformations and projectors specific to each.
While other, related, work has often used 3-D sections, our work natively
addresses the boundary polytopes and transformations in the Euclidean space ${\mathbi{G}^{\mathbi{6}}}$. 
The single point of view and the simple linear algebra involved makes the
presentation more consistent.

Some unexpected complexities have been encountered, such as the occurrence
of one boundary polytope that is the intersection of eight 5-D boundary polytopes, a fruitful
area for further investigation.

     %-------------------------------------------------------------------------
     % The back matter of the paper - acknowledgements and references
     %-------------------------------------------------------------------------

     % Acknowledgements come after the appendices

%\ack{
\section*{Acknowledgement}
~~\\
The authors acknowledge the invaluable assistance of Frances C. Bernstein.
\\
The work by Herbert J. Bernstein has been supported in part by NIH NIGMS grant 2R15GM078077-02.
The content is solely the responsibility of the authors and does not
necessarily represent the official views of the funding agency.

The authors wish to thank the referee who caught some errors in the dimensions assigned
to some of the boundary polytopes in an earlier draft of this paper.   The necessary recheck
produced one additional boundary polytope, and was the inspiration for the discussion
of the reduction in the number of degrees of freedom for some of the 2-folds and 3-folds.
%}

     % References are at the end of the document, between \begin{references}
     % and \end{references} tags. Each reference is in a \reference entry.

     %-------------------------------------------------------------------------
     % TABLES AND FIGURES SHOULD BE INSERTED AFTER THE MAIN BODY OF THE TEXT
     %-------------------------------------------------------------------------

\onecolumn

\begin{table}
\caption{Fifteen 5-D boundary polytopes on Niggli-reduced cells in ${\mathbi{G}^{\mathbi{6}}}$.  The special-position
subspaces are identified by the conditions to be added to the conditions that
define the boundary polytope itself.  For a given boundary polytope ${\mathitbf{\Gamma}}$, the
column ``Condition'' gives the ${\mathbi{G}^{\mathbi{6}}}$ constraints (prior to closure) of the
boundary polytope.  When taken with the ``Special-Position Subspace'' constraint
in the last column, the result is the entirety of the special-position subspace $\hat{{\mathitbf{\Gamma}}}$.
The ``Special-Position Subspace'' constraint by itself is ${\mathitbf{\Gamma}}^{\prime}$.   Boundary polytopes 1 and 2
apply in both the all acute ($+ + +$) and all obtuse  ($- - -$) branches of the Niggli-reduced
cone.  Boundary polytopes 8, B, E and F are restricted to the all obtuse ($- - -$) branch of the
Niggli-reduced cone, $\mathitbf{N}$.  Boundary polytopes 6, 7, 9, A, C and D  are restricted to the all acute ($+ + +$) branch of $\mathitbf{N}$.
While the boundary polytopes 3, 4 and 5 are boundaries of both the all acute ($+ + +$) and all obtuse  ($- - -$) branches, the common
special position subspace of those polytopes is just $g_4 = g_5 = g_6 = 0$ which is part of the ($- - -$) branch.}

\begin{center}
\begin{tabular}{|l|l|l|l|}

\hline
{\bf Class}& {\bf Case}& {\bf Condition}&{\bf Special-Position Subspace}\\
\hline
Equal cell edges& 1 &$g_1 = g_2$&$g_4 = g_5$\\
&2 &$g_2 = g_3$&$g_5 = g_6$\\
\hline
Ninety degrees&3 &$g_4 = 0$&$g_5 = g_6 = 0$\\
&4 &$g_5 = 0$&$g_4 = g_6 = 0$\\
&5 &$ g_6 = 0$&$g_4 = g_5 = 0$\\
\hline
Face diagonal
&6 &$g_2 = g_4 $ and $ g_5 \geqslant  g_6$&(none)\\
&7 &$g_2 = g_4 $ and $ g_5 <  g_6$&$g_5 = g_6/2$\\
&8 &$g_2 = -g_4$&(none)\\
&9 &$g_1 = g_5 $ and $ g_4 \geqslant  g_6$&(none)\\
&A &$g_1 = g_5 $ and $ g_4 < g_6$&$g_4 = g_6/2$\\
&B &$g_1 = -g_5$&(none)\\
&C &$g_1 =  g_6 $ and $ g_4 \geqslant g_5$&(none)\\
&D &$g_1 =  g_6 $ and $ g_4 < g_5$&$g_4 = g_5/2$\\
&E &$g_1 = - g_6$&(none)\\
\hline
Body diagonal &F &$g_1+g_2+g_3+g_4+g_5+ g_6 = g_3$
&$g_1-g_2-g_4+g_5=0$\\
\hline
\end{tabular}
\end{center}
\label{5D}
\end{table}%

\begin{table}
\caption{Roof/Niggli symbol, International Tables (IT) lattice character, Bravais lattice type, ${\mathbi{G}^{\mathbi{6}}}$ subspace \cite{Andrews1988}, ${\mathbi{G}^{\mathbi{6}}}$ boundary polytope.}
\begin{center}
\begin{tabular}{|c|c|c|c|c|}
\hline
{\bf Roof/}     &{\bf IT}&{\bf Bravais}&{\bf ${\mathbi{G}^{\mathbi{6}}}$}&{\bf ${\mathbi{G}^{\mathbi{6}}}$}\\
{\bf Niggli}&{\bf Lattice}&{\bf Lattice}&{\bf Subspace}&{\bf Boundary}\\
{\bf Symbol}&{\bf Char}&{\bf Type}&                             & {\bf Polytope} \\
\hline
44A&3&${\bf cP}$&$(r,r,r,0,0,0)$&$12345\!=\!12\hat{3}\!=\!12\hat{4}\!=\!12\hat{5}$\\
\hline
44C&1&$cF$&$(r,r,r,r,r,r)$&12679ACD\\
\hline
44B&5&$cI$&$(r,r,r,\!-\!2r/3,\!-\!2r/3,\!-\!2r/3)$&$\text{12F2}^{\prime} \text{F}^{\prime} = 1\hat{2}\hat{\text{F}}$\\
\hline
45A&11&${\bf tP}$&$(r,r,s,0,0,0)$&$1345 = 1\hat{3} = 1\hat{4} = 1\hat{5}$\\
45B&21&${\bf tP}$&$(r,s,s,0,0,0)$&$2345 = 2\hat{3} = 2\hat{4} = 2\hat{5}$\\
\hline
45D&6&$tI$&$(r,r,r,-r+s,-r+s,-2s)$&$\text{12FF}^{\prime} = 12\hat{\text{F}}$\\
45D&7&$tI$&$(r,r,r,-2s,-r+s,-r+s)$&$\text{12F2}^{\prime} = 1{\hat{2}}\text{F}$\\
45C&15&$tI$&$(r,r,s,-r,-r,0)$&158BF\\
45E&18&$tI$&$(r,s,s,r/2,r,r)$&$\text{2ADA}^{\prime}  = 2{\hat{A}}\text{D}$\\
\hline
48A&12&${\bf hP}$&$(r,r,s,0,0,-r)$&134E\\
48B&22&${\bf hP}$&$(r,s,s,-s,0,0)$&2458\\
\hline
49C&2&$hR$&$(r,r,r,s,s,s)$&$121^\prime 2^\prime = \hat{1}\hat{2}$\\
49D&4&$hR$&$(r,r,r,-s,-s,-s)$&$121^\prime 2^\prime = \hat{1}\hat{2}$\\
49B&9&$hR$&$(r,r,s,r,r,r)$&1679ACD\\
49E&24&$hR$&$(r,s,s,\!-s\!+\!r/3,\!-2r/3,\!-2r/3)$&$2F2^\prime \text{F}^\prime = \hat{2}\hat{\text{F}}$\\
\hline
50C&32&${\bf oP}$&$(r,s,t,0,0,0)$&$345 = \hat{3} = \hat{4} = \hat{5}$\\
\hline
50D&13&$oC$&$(r,r,s,0,0,-t)$&134\\
50E&23&$oC$&$(r,s,s,-t,0,0)$&245\\
50A&36&$oC$&$(r,s,t,0,-r,0)$&35B\\
50B&38&$oC$&$(r,s,t,0,0,-r)$&34E\\
50F&40&$oC$&$(r,s,t,-s,0,0)$&458\\
\hline
51A&16&$oF$&$(r,r,s,-t,-t,-2r+2t)$&$\text{1F1}^{\prime} = \hat{1}\text{F}$\\
51B&26&$oF$&$(r,s,t,r/2,r,r)$&$\text{ADA}^{\prime}  = \hat{A}\text{D}$\\
\hline
52A&8&$oI$&$(r,r,r,-s,-t,-2r+s+t)$&12F\\
52B&19&$oI$&$(r,s,s,t,r,r)$&29C = 2AD\\
52C&42&$oI$&$(r,s,t,-s,-r,0)$&58BF\\
\hline
\end{tabular}
\end{center}
\label{NiggliFormsI}
\end{table}%

\begin{table}
\caption{Roof/Niggli symbol, International Tables (IT) lattice character, Bravais lattice type, ${\mathbi{G}^{\mathbi{6}}}$ subspace, ${\mathbi{G}^{\mathbi{6}}}$ boundary polytope, continued}
\begin{center}
\begin{tabular}{|c|c|c|c|c|}
\hline
{\bf Roof/}     &{\bf IT}&{\bf Bravais}&{\bf ${\mathbi{G}^{\mathbi{6}}}$}&{\bf ${\mathbi{G}^{\mathbi{6}}}$}\\
{\bf Niggli}&{\bf Lattice}&{\bf Lattice}&{\bf Subspace}&{\bf Boundary}\\
{\bf Symbol}&{\bf Char}&{\bf Type}&                             & {\bf Polytope} \\
\hline
53A&33&${\bf mP}$&$(r,s,t,0,-u,0)$&35\\
53B&35&${\bf mP}$&$(r,s,t,-u,0,0)$&45\\
53C&34&${\bf mP}$&$(r,s,t,0,0,-u)$&34\\
\hline
55A&10&$mC$&$(r,r,s,t,t,u)$&$\text{11}^{\prime} = \hat{1}$\\
55A&14&$mC$&$(r,r,s,t,t,u)$&$\text{11}^{\prime} = \hat{1}$\\
57B&17&$mC$&$(r,r,s,-t,-u,-2r+t+u)$&1F\\
55B&20&$mC$&$(r,s,s,t,u,u)$&$\text{22}^{\prime} = \hat{2}$\\
55B&25&$mC$&$(r,s,s,t,u,u)$&$\text{22}^{\prime} = \hat{2}$\\
57C&27&$mC$&$(r,s,t,u,r,r)$&9C = AD\\
56A&28&$mC$&$(r,s,t,u,r,2u)$&$\text{AA}^{\prime} = \hat{A}$\\
56C&29&$mC$&$(r,s,t,u,2u,r)$&$\text{DD}^{\prime} = \hat{\text{D}}$\\
56B&30&$mC$&$(r,s,t,s,u,2u)$&$\text{77}^{\prime} =  \hat{7}$\\
54C&37&$mC$&$(r,s,t,-u,-r,0)$&5B\\
54A&39&$mC$&$(r,s,t,-u,0,-r)$&4E\\
54B&41&$mC$&$(r,s,t,-s,-u,0)$&58\\
57A&43&$mC$&$(r,s,t,-s+u,-r+u,-2u)$&$\text{FF}^{\prime} = \hat{\text{F}}$\\
\hline
\end{tabular}
\end{center}
\label{NiggliFormsII}
\end{table}%

\begin{table}
\caption{The Hosoya boundary polytopes of the three cubic lattice types and the primitive
hexagonal in terms of ${\mathbi{G}^{\mathbi{6}}}$}
\begin{center}
\begin{tabular}{|c|c|}
\hline
$Hosoya_1$&$2g_1 = g_5+g_6$\\
$Hosoya_2$&$2g_2 = g_4+g_6$\\
$Hosoya_3$&$2g_3 = g_4+g_5$\\
$Hosoya_4$&$g_4 = g_5$, the special-position subspace of case 1\\
$Hosoya_5$&$g_5 = g_6$, the special-position subspace of case 2\\
$Hosoya_6$&$g_4 = g_5/2$, the special-position subspace of case D \\
\hline
\end{tabular}
\end{center}
\label{hosoyaboundaries}
\end{table}%

\bibliographystyle{plain}
\bibliography{G6_boundaries_2Aug13}

%\referencelist[G6_boundaries_2Aug13]

\appendix
\section{{\bf Supplementary Materials -- The Monte Carlo Search for Boundary Polytopes}}

\label{sec:mc}

One can follow an entirely algebraic process to identify all boundary polytopes.  In view of the complexity
of the space it is helpful to,
use a computer to explore ${\mathbi{N}}$ especially to confirm information on the dimensionality of boundary polytopes
and on the necessary boundary transformations.

In order to reduce the computational burden and deal
with the open boundary polytopes, rather than randomly generating points in ${\mathbi{G}^{\mathbi{6}}}$, we generate random short 
lines in ${\mathbi{G}^{\mathbi{6}}}$ and look for cases in which the boundary of ${\mathbi{N}}$ must have been crossed because one end of the line is Niggli-reduced, while the other end is not.  The process for the initial search for 5-D boundary polytopes is given in Table \ref{randsearch1}.

\begin{figure} 
\caption{Illustration of Monte Carlo search for boundary polytopes in ${\mathbi{G}^{\mathbi{6}}}$.  The hatched line represents a boundary
polytope, {\it e.g.} $g_2 = g_4$.  The unhatched side of the boundary (${\mathbi{N}}$) consists of Niggli-reduced cells.
In this case the hatched side is divided into two regions,  ${\mathbi{N\!N1}}$ ({\it e.g.} $g_5 \geqslant g_6$) and 
${\mathbi{N\!N2}}$ ({\it e.g.} $g_5 < g_6$) for which different reduction matrices are required to get a 
Niggli-reduced cell, {\it e.g.} $M_6$ for  ${\mathbi{N\!N1}}$ and $M_7$ for ${\mathbi{N\!N2}}$.  ``${\mathbi{NN}}$'' stands for not Niggli-reduced. 
 $V1$ is a randomly generated
probe point in region ${\mathbi{N}}$ for which a random short line reaches across the boundary to reach 
region ${\mathbi{N\!N1}}$, so the starting point is associated with $M_6$.  $V2$, $V4$ and $V5$  are randomly 
generated probe points for which the random short line remains in the Niggli-reduced
region ${\mathbi{N}}$.  Therefore $V2$, $V4$ and $V5$ are discarded.    $V3$ is a randomly generated
probe point in region ${\mathbi{N}}$ for which a random short line reaches across the boundary to reach 
region ${\mathbi{N\!N2}}$, so the starting point is associated with $M_7$.  Because of this difference in reduction
matrices, the boundary polytope is treated as consisting of two distinct boundary polytopes, in this
case, cases 6 and 7.} 
\scalebox{.5}{\includegraphics{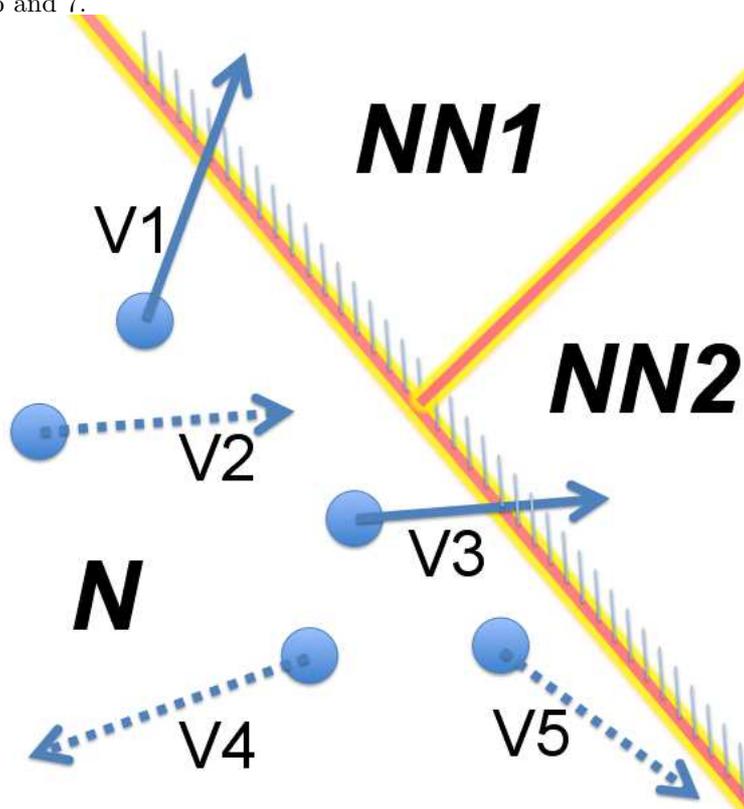}} 
\label{fig:boundary_probes}
\end{figure}

\begin{figure} 
\caption{Counts of points found near various boundary polytopes in 100 million trials, organized in declining
order of counts, showing the most populated 23 of the 92 boundary polytopes found in this run.  This is a run
with no filtering for any particular boundary with the counts shown on a logarithmic scale.  Note
the precipitous drop of nearly 2 orders of magnitude after the first 15 boundary polytopes.  This drop confirms
that those 15 boundary polytopes are the 5-D boundary polytopes and that there is a vanishingly small probability of there
being any other 5-D boundary polytopes of the Niggli-reduced cells.} 
\scalebox{.6}{\includegraphics{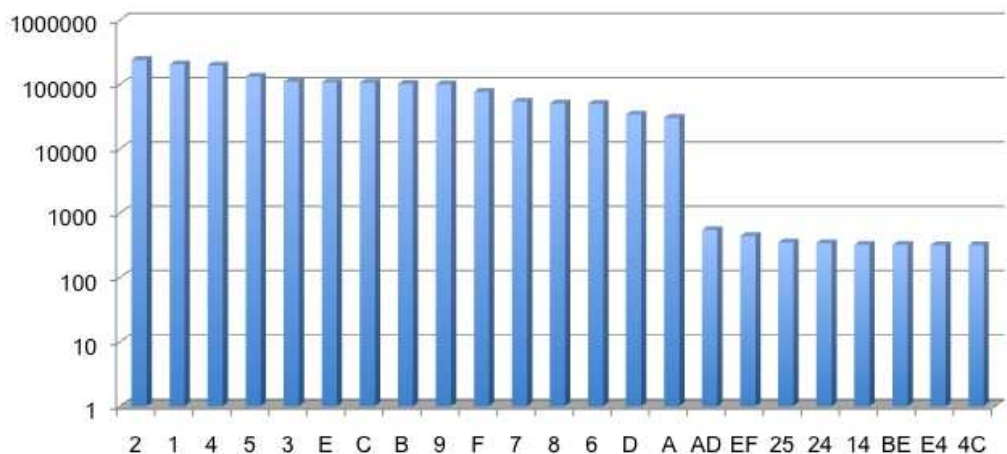}}
\label{fig:boundary_counts}
\end{figure}

The search for the boundary polytopes resulting from the process in Table \ref{randsearch1} produces
transformations in the course of Niggli reduction (step 6).   We sort the Niggli-reduced ${\mathbi{G}^{\mathbi{6}}}$ vectors
by the associated transformation matrix from step 6.  A high population for a given matrix indicates a
significant volume of ${\mathbi{G}^{\mathbi{6}}}$ with access to the associated boundary polytope, implying a 5-D boundary polytope
(See Fig. \ref{fig:boundary_probes}  and Fig. \ref{fig:boundary_counts}).
A lower population for a given matrix possibly implies a lower-dimensional boundary polytope, a common ``edge'' resulting from
the intersection of multiple 5-D boundary polytopes.    With enough probes, many of the lower
dimensional boundary polytopes can be discovered, but only with difficulty because the information
on those lower-dimensional boundary polytopes is swamped in a sea of data about the 5-D boundary polytopes.

Efficient discovery of all these intersections of multiple 5-D boundary polytopes
requires the use of ``\textbf{projectors}''.  A projector for a subspace ${\mathbi{Q}}$  in ${\mathbi{G}^{\mathbi{6}}}$ is a symmetric matrix, $P$, that when
applied to any point $\overrightarrow{g}$ in ${\mathbi{G}^{\mathbi{6}}}$  is such that $P\overrightarrow{g}$ is the closest point in ${\mathbi{Q}}$ to $\overrightarrow{g}$.  A projector has
the nice property that $PP = P$, i.e. that it acts like the identity matrix on the subspace ${\mathbi{Q}}$.  

The projector for a given boundary polytope can be discovered by examining the set of Niggli-reduced
${\mathbi{G}^{\mathbi{6}}}$ vectors associated with the transform matrix for that boundary polytope.  The examination can
either be a simple inspection of the list of vectors or can be done algorithmically by use of
a singular value decomposition (SVD) calculation \cite{Beltrami1873,Jordan1874,Stewart1993}.
It should be noted that the projector projects onto the hyperplane associated with the polytope and
may project a point from ${\mathbi{G}^{\mathbi{6}}}$, or even from ${\mathbi{N}}$ itself, outside of ${\mathbi{N}}$.

Let us first derive projectors by simple inspection.  A search for points near a boundary produces a list of vectors that can be examined for the conditions that are met.  For instance, the trial vectors:

\begin{center}
\begin{tabular}{llllll}
4.41605 &{\bf 53.21164} &{\bf 53.3171}  &-9.85206   &-2.73956   &-1.78806\\
4.95245 &{\bf 106.2402} &{\bf 106.5968}  &-72.3608  &-0.26549   &-4.79911\\
5.62821 &{\bf 98.26772} &{\bf 98.36612}     &24.37056   &1.57819    &1.85157\\
\end{tabular}
\end{center}

\noindent{}are seen to meet the condition $g_2=g_3$.

The trial vectors:

\begin{center}
\begin{tabular}{llllll}
4.85822 &9.79018    &40.14963   &-3.6758    &{\bf -0.01092} &-2.18456\\
4.89205 &21.22063   &75.92303   &6.01777&{\bf 0.05752}  &2.78514\\
5.03365 &26.32789   &46.84058   &-17.2646   &{\bf -0.01252} &-2.03757\\
\end{tabular}
\end{center}

\noindent{}are seen to meet the condition $g_5=0$.

In each case the necessary projector is obvious (see projectors $P_2$ and $P_3$, below).   In
other cases, such as the body-diagonal boundary polytope, or many of the lower-dimensional
boundary polytopes involving multiple face-diagonals, simple inspection is challenging.
Given a sufficient number of vectors, the projector may be recovered algorithmically from such
a list of vectors, rather than by inspection.  If a singular value decomposition is
done on a list of vectors near to and covering  a single 5-D boundary polytope, the vector corresponding to 
the smallest singular value is the unit normal to that boundary polytope, and the projector 
onto that polytope is simply the identity minus the projector onto the line of the normal.  The projector
onto the line of the unit normal is generated by forming a matrix, $A$,
 whose first row is the unit normal and whose remaining rows are set to zero.  Then $A^T\!A$ is the projector onto the line of that unit normal.
 
In the general case of a lower-dimensional boundary polytope, the projector onto the boundary polytope
is the identity minus the projector $Q$ onto the hyperplane ${\mathitbf{\Omega}}$ orthogonal  to the boundary
polytope.  In that case a singular value decomposition on the list of vectors near to and covering
the lower-dimensional boundary polytope will have small singular values for the vectors spanning ${\mathitbf{\Omega}}$.
The projector $Q$ onto ${\mathitbf{\Omega}}$ is generated by forming a matrix $A$ with initial rows consisting of
the vectors corresponding to the small singular values and the remaining rows set to zero. 
Then $A^T\!A$ is the projector onto ${\mathitbf{\Omega}}$ and $I- A^T\!A$ is the projector onto the boundary
polytope.

Once we have found the projectors for the 5-D boundary polytopes, the projectors for their intersections
may be generated from their products.  The product of two different projectors is not, itself, likely to
be a projector, but repeated squaring of that product rapidly converges to the correct
projector \cite{Andrews1976}.  Given these projectors, the process for the remaining
boundary polytopes is given in Table \ref{randsearch2}.  This process is similar to the initial
process, but three new steps have been added:  In step 0 the projector for a particular boundary polytope is read in. 
In step 2A the vector confirmed by step 2 is projected onto the hyperplane containing the boundary polytope, and step 2B verifies that the projected vector is still a valid unit cell.
When doing a random perturbation of a vector already projected onto the hyperplane containing a boundary polytope while looking for intersections with that boundary polytope,
it is sufficient to use a random vector projected onto that the hyperplane containing the boundary polytope.  However, that risks missing some very low dimension cases.
That search produced 215 distinct boundary polytopes.  The search was then redone, stepping back into the Niggli cone after
each projection, and using a full 6-D spherical random search to ensure catching any nearby boundary polytopes.  That 
search added one more distinct boundary polytope, for a total of 216 boundary polytopes.

The dimension of a boundary polytope can be determined by computing the number of eigenvalues equal
to 1 of the projector onto the hyperplane containing the boundary polytope, and having distinct boundary 
projectors is a sufficient condition  for two
boundary polytopes to be distinct.  However, 
distinct projectors is a necessary, but not sufficient, condition for two boundary polytopes
to be distinct, because crossing one bounding hyperplane in two
different places can require two different transformation matrices to reduce the result.  The reduction transformation
matrices themselves are needed to disambiguate cases with the identical bounding hyperplane projector.

In the course of the Monte Carlo investigation of the  boundary polytopes resulting from combinations of
5-D boundary polytopes, we track the relative populations as an indicator of a consistent assignment
of dimension.  The candidates for assignment to a particular dimension are sorted by population,
and the list is cut off on a precipitous drop of more than an order of magnitude.  For the remaining
populations, the mean population $\mu$ and estimated standard deviation $\sigma$ are
computed.  For each population $\tau$, a Z-score, $(\tau - \mu)/\sigma$, is computed 
(see http://en.wikipedia.org/wiki/Standard\_score).  Z-scores
of less than -1 suggest the possible need for further study of the boundary polytopes involved.   A combination
of the population analysis resulting from the Monte Carlo search, algebraic analysis of constraints and computation of the eigenvalues of 
projectors allows the identification of the dimensions of the boundary polytopes, helps to identify
distinct combinations of intersections of 5-D boundary polytopes that represent the same lower dimensional boundary polytopes
and confirms the critical initial identification of the 15 5-D boundary polytopes identified purely by the Monte Carlo search.

We limit our consideration of valid boundary polytopes to those avoiding the mathematically interesting but crystallographically impossible cases of zero length cell edges.    Combinations of boundary polytopes without a valid intersection or with an intersection that would force any of $g_{\{1,2,3\}}$ to zero or that  
did not have neighboring Niggli-reduced probe points are eliminated.  574 combinations of 1 through 8 intersecting 5-D boundary
polytopes were not degenerate.    Many combinations represent equivalent boundary polytopes.   There are 216 distinct boundary 
polytopes.  There are
15 5-D boundary polytopes of the full ${\mathbi{G}^{\mathbi{6}}}$ Niggli cone,  53 4-D boundary polytopes resulting from intersections of pairs of the 15 5-D boundary polytopes, 79 3-D boundary polytopes resulting from the 2-fold and higher intersections of the 15 5-D boundary polytopes, 55 2-D boundary polytopes resulting from 2-fold and higher intersections  of the 15 5-D boundary polytopes, 14 1-D boundary polytopes resulting from 3-fold and higher intersections of the 15 5-D  boundary polytopes.  The ability of the intersection of only 2 5-D boundary polytopes
to produce 4-D, 3-D or 2-D boundary polytopes results from the additional constraints imposed by Niggli reduction.

For the 5-D boundary polytopes, three special cases arise in which the Niggli reduction conditions divide a single polytope into two sub-polytopes, the ``flat boundary intersections''.
In each those three cases, instead of two boundary polytopes meeting at an angle,  two boundary polytopes meet edge on.  These three cases are $g_4=g_2$, $g_5=g_1$,  and $g_6=g_1$.  In each case the division in the polytope is based on the equality of the other two members of $g_{\{4,5,6\}}$.  For example, when $g_4=g_2$, the division is along $g_5 = g_6$.  Therefore, when multiplying projectors, instead of multiplying the projector for $g_4=g_2$ by itself when the two half-polytopes are required, the second projector is replaced by the projector for the division, \textit{i.e.} in the example given instead of multiplying the projector for $g_4=g_2$ by itself, it is multiplied by the projector for $g_5 = g_6$.

\begin{table}
\caption{Initial process to locate 5-D boundary polytopes}
\begin{center}
\begin{tabular}{lp{12.5cm}}

1. &Generate a random vector in ${\mathbi{G}^{\mathbi{6}}}$.\\
2. &Confirm that the vector represents a proper unit cell (for instance, the sum of the interaxial angles must be less than 360 degrees).\\
3. &Niggli-reduce the vector.\\
4. &Randomly perturb the vector resulting from step 3.\\
5. &Confirm that the perturbed vector represents a proper unit cell.\\
6. &Niggli-reduce the vector from step 5, accumulating the total transformation matrix from the vector of step 3 to the new reduced vector.\\
7. &If the transformation is the unit matrix, then the perturbed vector was not near the boundary of ${\mathbi{N}}$. Discard the trial. Otherwise, proceed.\\
8. &If the transformation matrix has been discovered before, increment a counter for its occurrences. Otherwise, add it to the list of discovered transformations.\\
9. &Repeat many times, starting at step 1.\\
\end{tabular}
\end{center}
\label{randsearch1}
\end{table}%

\begin{table}
\caption{Process to locate lower-dimensional boundary polytopes.}
\begin{center}
\begin{tabular}{lp{12.5cm}}

0.  &{\bf Read in the projector for the desired boundary to be probed.}\\
1. &Generate a random vector in ${\mathbi{G}^{\mathbi{6}}}$.\\
2. &Confirm that the vector represents a proper unit cell (for instance, the sum of the interaxial angles must be less than 360 degrees).\\
2A. &{\bf Project the vector confirmed in step 2 onto the boundary.}\\
2B. &{\bf Confirm that the projected vector is still a valid unit cell. }\\
3. &Niggli-reduce the vector.\\
4. &Randomly perturb the vector resulting from step 3.\\
5. &Confirm that the perturbed vector represents a proper unit cell.\\
6. &Niggli-reduce the vector from step 5, accumulating the total transformation matrix from the vector of step 3 to the new reduced vector.\\
7. &If the transformation is the unit matrix, then the perturbed vector was not near the boundary of ${\mathbi{N}}$. Discard the trial. Otherwise, proceed.\\
8. &If the transformation matrix has been discovered before, increment a counter for its occurrences. Otherwise, add it to the list of discovered transformations.\\
9. &Repeat many times, starting at step 1.\\
\end{tabular}
\end{center}
\label{randsearch2}
\end{table}%

\begin{table}
\caption{The 53 4-D boundary polytopes.  The boundary conditions use the symbol notation of Table \ref{5D}
for the 5-D boundary polytopes.  The bounding 4-D polytopes of each of the 15 5-D boundary polytopes can be read
off from this table by going down the appropriate column to the main diagonal and then across.  For example, the edges of
the 3 polytope are 13, 23, 34, 35, 3A, 3B, 3D and 3E.}
\begin{center}
\begin{tabular}{|c|ccc|ccc|ccc|ccc|c|}
\hline
$g_2\!=\!g_3$&$g_4\!=\!0$&$g_5\!=\!0$&$g_6\!=0$&
\multicolumn{3}{|c|}{a-face-diagonals}&\multicolumn{3}{|c|}{b-face-diagonals}&\multicolumn{3}{|c|}{c-face-diagonals}&body\\
\hline
12 & 13 & 14 & 15 & 16 &17 &18 &19 &1A &1B  &1C &1D &1E &1F \\
      & 23 & 24 & 25 & 26 & 27 & 28 &29 & 2A & 2B & 2C & 2D &2E &2F \\
\hline
      &      & 34 &35  &       &       &      &      & 3A & 3B &        &3D & 3E &\\
     &       &       &45 &       & 47  & 48 &     &       &       & 4C  &      & 4E &\\
     &      &        &      & 56 &        & 58 & 59&      & 5B  & & & &\\
\hline
     &     &        &       &       &67   &       & 69&      &   & & & &\\
     &     &        &       &       &       &     &        &       &       &7C& & & \\
    &     &        &       &       &       &     &        &       &       &     &       &       &8F\\
\hline
      &     &        &       &       &       &     &        &9A &      &9C& & &\\
    &     &        &       &       &       &     &        &       &       &     &AD & &\\
    &     &        &       &       &       &     &        &       &       &     &       & BE &BF\\
\hline
    &     &        &       &       &       &     &        &       &       &     &CD& &\\
    &     &        &       &       &       &     &        &       &       &     &       &        &EF\\
\hline
\end{tabular}
\end{center}
\label{4D}
\end{table}%

\begin{table}
\caption{The 79 3-D boundary polytopes.  In some cases the most natural presentation
of a given 3-D polytope is a 4-fold intersection.   In each of those case an equivalent 3-fold
intersection is given in parentheses immediately below the 4-fold.}
\begin{center}
\begin{tabular}{|ccc|ccc|ccc|ccc|c|}
\hline
$g_4\!=\!0$&$g_5\!=\!0$&$g_6\!=\!0$&
\multicolumn{3}{|c|}{a-face-diagonals}&\multicolumn{3}{|c|}{b-face-diagonals}&\multicolumn{3}{|c|}{c-face-diagonals}&body\\
\hline
123 &124 &125 & 126 &127 & 128 &129 &12A &12B  &12C  & 12D & 12E & 12F\\
        &134 &135 &         &        &        &        & 13A & 13B &          & 13D & 13E &       \\
         &         &145 &         &147 &148 &        &          &         & 14C &          & 14E &       \\
        &        &         &156 &         &158 &159 &         &15B  &          &          &         &\\
        &        &         &        &167  &        &169 &         &         &          &           &         &\\
        &        &         &        &         &        &        &         &         &  17C &            &         &\\    
        &        &         &        &         &       &         &         &        &          &            &        & 18F \\
        &       &        &         &        &         &        &  19A  &        &         &          &         &\\
        &       &        &         &        &         &        &          &         &        & 1AD &         &\\
        &        &         &        &         &       &         &         &        &          &            &        &1BF\\
        &        &         &        &         &       &         &         &        &          &1CD   &        &\\
        &        &         &        &         &       &         &         &        &          &            &        &1EF \\
\hline
        &234 &235 &         &           &       &         &23A & 23B &        &23D    & 23E& \\
        &        &245 &        & 247   &248 &         &         &        & 24C &           & 24E& \\
        &        &        & 256&           &258 & 259 &         & 25B &        &         &           & \\
        &        &        &        &267&           & 269 &         &          &         &        &          &\\
        &        &        &         &        &         &        &        &         &  27C  &         &         &\\
        &        &        &        &          &        &       &         &       &        &         &           &28F \\
        &        &        &        &          &        &       &  29A  &         & 29C &         &         &\\
        &        &        &        &          &        &       &           &        &       &2AD   &         &\\
        &        &        &        &          &        &       &           &        &       &           &2BE &2BF  \\
        &        &        &        &          &        &       &           &        &       &2CD &         &\\
        &        &        &        &          &        &       &           &        &       &         &         &2EF \\
\hline
        &        &345&        &          &        &       &           &        & 34CD &         & 34E &\\
        &        &      &        &          &        &       &           &        & (34C)&         & &\\
        &        &        &        &          &        &359A&           &35B&       &         &         & \\
        &        &        &        &          &        &(359)&           &        &       &         &         & \\
        &        &          &        &       &           &        &       &         &      &3AD &         & \\
        &        &          &        &       &           &        &       &         &      &         &3BE   &\\
\hline
        &        &        &         &4567&         &         &         &         &          &           &         &\\
       &        &        &         &(456)&         &         &         &         &          &           &         &\\
        &        &        &         &        &458   &         &         &         &          &           &         &\\
        &        &        &         &        &         &        &        &         & 47C  &            &48EF  &\\
        &        &        &         &        &         &        &        &         &          &            &(48E) &\\
\hline
        &        &        &        &          &        & 569&         &           &          &            &        &\\
        &        &        &        &          &        &    &         &58BF &          &            &        &\\
        &        &        &        &          &        &    &         &(58B)&          &            &        &\\
\hline
        &        &        &        &          &        &679C&         &     &          &            &        &\\
        &        &        &        &          &        &(679)&         &     &          &            &        &\\
\hline
        &        &        &        &          &        &          &         &     &9ACD&            &        &\\
        &        &        &        &          &        &          &         &     &(9AC)&            &        &\\
\hline
       &        &         &        &         &       &         &         &        &          &            &        &BEF\\
\hline
\end{tabular}
\end{center}
\label{3D}
\end{table}%

\begin{table}
\caption{The 55 2-D boundary polytopes}
\begin{center}
\begin{tabular}{|cc|ccc|ccc|ccc|c|}
\hline
$g_5\!=\!0$&$g_6\!=\!0$&
\multicolumn{3}{|c|}{a-face-diagonals}&\multicolumn{3}{|c|}{b-face-diagonals}&\multicolumn{3}{|c|}{c-face-diagonals}&body\\
\hline
1234 &1235 &          &         &         &         &123A&123B&         &123D&123E&\\
          &1245 &          &1247&1248&         &         &         &124C&          &124E&\\
          &          &1256 &         &1258&1259&         &125B&         &          &          &\\
          &          &          &1267&         &1269&         &          &127C&        &           &\\
          &          &         &          &        &          &         &          &         &         &           &128F\\
          &          &         &          &        &          &129A&          &         &         &           &\\
          &          &         &          &        &          &         &          &         &12AD&          &\\
          &          &         &          &        &          &         &          &         &         &           &12BF\\
          &          &         &          &        &          &         &          &         &12CD&           &\\
          &          &         &          &        &          &         &          &         &         &           &12EF\\
 \hline
          &1345&         &          &        &           \multicolumn{3}{r|}{134CD} &         &         &134E&\\
          &         &         &          &        &          \multicolumn{3}{r|}{(134C)}    &         &  & &\\
          &         &         &          &        &          &         &             &         &         &  &\\
          &          &         &          &        & \multicolumn{2}{r}{1359A} &135B&         &          &          &\\
          &          &         &          &        & \multicolumn{2}{r}{(1359)} &         &         &          &          &\\
         &          &         &          &         &          &         &          &         &13AD&13BEF&\\
         &          &         &          &         &          &         &          &         &          &(13BE)&\\
 \hline
          &          &14567&          &1458&          &         &          &         &         &          &\\
          &          &(1456)&          &        &          &         &          &         &         &          &\\
          &          &         &          &        &          &         &          &147C &         &          &\\
          &          &         &          &        &          &         &          &           &148EF &          &\\
          &          &         &          &        &          &         &          &           &(148E) &          &\\
\hline
          &          &         &          &        &1569&         &          &           &         &          &\\
         &          &         &          &        &          &         &158BF&           &         &          &\\
         &          &         &          &        &          &         &(158B)&           &         &          &\\
\hline
         &          &         &          &       &          &          &           &\multicolumn{3}{l|}{1679ACD}&\\
          &          &         &          &        &          &         &           &(6D, 7A)&         &          &\\

\hline
          &2345&         &          &        &          &         &          &234CD&         &234E&\\
          &         &         &          &        &          &         &          &(234C)  &         &         &\\
          &          &         &          &        &\multicolumn{2}{l}{2359A} &235B&           &         &          &\\
          &          &         &          &        &\multicolumn{2}{l}{(2359)} &         &           &         &          &\\
          &          &         &          &        &          &         &          &         &23AD&          &\\
          &          &         &          &        &          &         &          &         &         &23BE&\\
\hline
          &          &\multicolumn{2}{l}{24567}          &2458 &          &         &          &         &         &           &\\
          &          &\multicolumn{2}{l}{(2456)}        &            &          &         &          &         &         &           &\\
          &          &         &          &        &          &         &          & 247C&         &          &\\
          &          &         &          &        &          &         &          &\multicolumn{3}{l|}{248EF}&\\
          &          &         &          &        &          &         &          &\multicolumn{3}{l|}{(248E)}&\\
\hline
          &          &         &          &        &2569 &         &          &         &         &           &\\
          &          &         &          &        &\multicolumn{3}{r|}{258BF}&         &         &           &\\
          &          &         &          &        &\multicolumn{3}{r|}{(258B)}&         &         &           &\\
\hline
          &          &         &          &        &\multicolumn{3}{c|}{2679C}&         &         &           &\\
          &          &         &          &        &\multicolumn{3}{c|}{(2679)}&          &         &          &\\
\hline
          &          &         &          &        &\multicolumn{3}{r|}{29ACD}&         &         &           &\\
          &          &         &          &        &\multicolumn{3}{r|}{(29AC)}&         &         &           &\\
\hline
          &          &         &          &        &          &         &          &         &         &           &2BEF\\
\hline
\end{tabular}
\end{center}
\label{2D}
\end{table}%

\end{document}